\global\def\draftcontrol{0}

   \def\versionno{  del Pezzo, Markov, and Seiberg  } 

\catcode`\@=11 

\expandafter\ifx\csname draftcontrol\endcsname\relax\global\def\draftcontrol{0} 
\fi 

{\count255=\time\divide\count255 by 60 
\xdef\hourmin{\number\count255} 
\multiply\count255 by-60\advance\count255 by\time 
\xdef\hourmin{\hourmin:\ifnum\count255<10 0\fi\the\count255}} 
\def\draftdate{\number\month/\number\day/\number\year\ \ \ \hourmin } 

\newcommand\makepapertitle{\par

  \begingroup 
    \renewcommand\thefootnote{\@fnsymbol\c@footnote}%
    \def\@makefnmark{\rlap{\@textsuperscript{\normalfont\@thefnmark}}}%
    \long\def\@makefntext##1{\parindent 1em\noindent 
            \hb@xt@1.8em{%
                \hss\@textsuperscript{\normalfont\@thefnmark}}##1}%
     \newpage 
     \global\@topnum\z@   
     \@makepapertitle 
     \thispagestyle{empty}\@thanks 
  \endgroup 
  \setcounter{footnote}{0}%
  \global\let\thanks\relax 
  \global\let\makepapertitle\relax 
  \global\let\@makepapertitle\relax 
  \global\let\@thanks\@empty 
  \global\let\@author\@empty 
  \global\let\@date\@empty 
  \global\let\@title\@empty 
  \global\let\title\relax 
  \global\let\author\relax 
  \global\let\date\relax 
  \global\let\and\relax 
  \def\version{\let\version\@version\@gobble} 
} 
\def\@makepapertitle{%
  \newpage 
   \ifnum\draftcontrol=1 {} 
   \version\versionno 
   \vskip 5em%
   \else 
   \hfill\hbox to 3cm {\parbox{4cm}{\@pubnum}\hss}%
   \vskip 5em%
   \fi 
   \begin{center}%
   \let \footnote \thanks 
      {\hskip -0\textwidth \hbox to 1\textwidth%
        {\centerline{\Large\bf{\noindent\@title}}}}%
     \vskip 1.5em%
     {\normalsize
       \lineskip .5em%
       \begin{tabular}[t]{c}%
         \@author 
       \end{tabular}\par}%
     \vskip 1em%
     {\@bstract}%
     \end{center}%
     \vfill
     \@date%
     \vskip 1.5em%
     \noindent
     \rule{12em}{.02em}\par\noindent
     \@email%
   \par 
} 

\gdef\@pubnum{} 
\def\pubnum#1{%
  \gdef\@pubnum{#1}} 

\gdef\@bstract{} 
\def\Abstract#1{%
  \gdef\@bstract{%
   \parbox{\textwidth-0pc}{%
   \centerline{\bf Abstract}\penalty1000 
   \noindent
   \renewcommand\baselinestretch{1.0} 
   {#1}}} 
} 

\gdef\@email{}
\def\email#1{%
   \gdef\@email{%
   Email: {\tt #1}}
}

\def\ps@paper{\let\@mkboth\@gobbletwo%
     \ifnum\draftcontrol=1 
        \def\@oddfoot{\hbox to \textwidth{\tiny \versionno \hfil\tiny\draftdate}%
        \hskip -\textwidth \hbox to \textwidth{\hfil\rm\thepage\hfil}}%
     \else\def\@oddfoot{\hbox to \textwidth{\hfil\rm\thepage\hfil}} 
     \fi 
     \let\@evenfoot\@oddfoot 
} 

\def\body{\clearpage 
          \pagestyle{paper} 
        } 
\newenvironment{acknowledgments}{%
\vskip 3.25ex 
\noindent {\bf Acknowledgments} 
} 

\def\@version#1{\ifnum\draftcontrol=1 
\typeout{}\typeout{#1}\typeout{} 
\vskip3mm\centerline{\hbox{\fbox{\normalsize{\tt DRAFT -- #1 -- } 
                   {\draftdate}}}}\vskip3mm 
\fi} 
\let\version\@version 
\long\def\eqlabel#1{\ifnum\draftcontrol=1 
                    \tag@false  
                    \tag*{(\theequation) \hbox to -0.2cm{\hspace{0cm}\small{#1}\hss}} 
                    \refstepcounter{equation}  
                    \edef\@currentlabel{\theequation} 
                    \ltx@label{#1}          
                    \else 
                    \label{#1} 
                    \fi 
                    } 
\let\st@bibitem\@bibitem 
\let\st@lbibitem\@lbibitem 
\ifnum\draftcontrol=1 
  \def\@bibitem#1{%
    \st@bibitem{#1}\a@@label{#1}\ignorespaces} 
  \def\@lbibitem[#1]#2{%
    \st@lbibitem[#1]{#2}\a@@label{#2}\ignorespaces} 
  \def\a@@label#1{%
    \gdef\a@lab{\smash{\normalfont\small#1}} 
    \ifvmode 
      \if@inlabel 
        \global\setbox\@labels\hbox{%
          \llap{\a@lab\let\a@lab\relax 
                \kern\@totalleftmargin\kern\marginparsep}%
          \box\@labels}%
      \fi 
    \fi} 
\fi 

\documentclass[12pt,letterpaper]{article} 

\usepackage{amsmath,bm,amsfonts,amssymb,array,calc,amsthm,rotating,epsfig,psfrag} 
\usepackage[nosort]{cite} 
\usepackage{graphicx}

\renewcommand\baselinestretch{1.25} 
\renewcommand\section{\@startsection {section}{1}{\z@}%
                                   {-3.5ex \@plus -1ex \@minus -.2ex}%
                                   {2.3ex \@plus.2ex}%
                                   {\normalfont\large\bfseries}} 
\renewcommand\subsection{\@startsection{subsection}{2}{\z@}%
                                   {-3.25ex\@plus -1ex \@minus -.2ex}%
                                   {1.5ex \@plus .2ex}%
                                   {\normalfont\normalsize\bfseries}} 
\renewcommand\subsubsection{\@startsection{subsubsection}{3}{\z@}%
                                   {-3.25ex\@plus -1ex \@minus -.2ex}%
                                   {1.5ex \@plus .2ex}%
                                   {\normalfont\normalsize\it}} 
\renewcommand\paragraph{\@startsection{paragraph}{4}{\z@}%
                                   {-3.25ex\@plus -1ex \@minus -.2ex}%
                                   {1.5ex \@plus .2ex}%
                                   {\normalfont\normalsize\bf}} 
\renewcommand\subparagraph{\@startsection{subparagraph}{5}{\z@}%
                                   {-1.25ex\@plus -1ex \@minus -.2ex}%
                                   {0ex \@plus .2ex}%
                                   {\normalfont\normalsize\it}} 

\def\ie{{\it i.e.}} 
\def\eg{{\it e.g.}} 

\def\revise#1       {\raisebox{-0em}{\rule{3pt}{1em}}%
                     \marginpar{\raisebox{.5em}{\vrule width3pt\ 
                     \vrule width0pt height 0pt depth0.5em 
                     \hbox to 0cm{\hspace{0cm}{%
                     \parbox[t]{4em}{\raggedright\footnotesize{#1}}}\hss}}}}

\def\cala         {{\cal A}}

\def\cale         {{\cal E}} 
\def\calf         {{\cal F}} 
\def\calg         {{\cal G}} 
 
\def\calh         {{\cal H}} 
\def\cali         {{\cal I}}

\def\caln         {{\cal N}} 
\def\calo         {{\cal O}}

\def\complex      {{\mathbb C}} 
 
\def\projective   {{\mathbb P}}

\def\zet          {{\mathbb Z}}

\def\EE           {{\it e}} 
\def\ii           {{\it i}} 
 
\def\tr           {\mathop{\rm tr}}

\def\sqr#1#2{{\vcenter{\vbox{\hrule height.#2pt   
 \hbox{\vrule width.#2pt height#1pt \kern#1pt 
 \vrule width.#2pt}\hrule height.#2pt}}}}

\def\U{{\it U}}
\def\SU{{\it SU}}
\def\CZ{{\complex^3/\zet_3}}

\newcommand{\CP}[1]{\mathbb{P}^{#1}}

\newcommand{\R}[1]{\mathbb{R}^{#1}}

\def\RP{\mathbb{RP}}

\def\Z{\mathbb{Z}}

\def\Y{\mathbf{Y}}
\def\S{\mathbf{S}}
\def\T11{{T}^{1,1}}
\def\be{\begin{equation}}
\def\ee{\end{equation}}
\def\bear{\begin{eqnarray}}
\def\eear{\end{eqnarray}}

\def\ie{{\it i.e.\/}}

\def\bV{\mathbf{V}}

\def\bX{\mathbf{X}}

\def\dim{\mathrm{dim} \, }
\def\Vol{\mathrm{Vol}}

\def\mO{\mathcal{O}}

\def\tr{\mathrm{tr}\,}

\def\bi{\bibitem}
\def\ch{\mathrm{ch}}
\def\Td{\mathrm{Td}}
\def\Ext{\mathrm{Ext}}
\def\gcd{\mathrm{gcd}}
\def\lcm{\mathrm{lcm}}
\def\Hom{\mathrm{Hom}}
\def\deg{\mathrm{deg}}

\catcode`\@=12 

\begin{document} 


\title{Dibaryons from Exceptional Collections}

\pubnum{%
NSF-KITP-03-50 \\ 
hep-th/0306298} 
\date{June 2003} 

\author{Christopher P. Herzog and Johannes Walcher \\[0.4cm] 
\it Kavli Institute for Theoretical Physics \\ 
\it University of California \\ 
\it Santa Barbara, CA 93106, USA \\[0.2cm] 
}

\email{herzog@kitp.ucsb.edu, walcher@kitp.ucsb.edu}

\Abstract{
We discuss aspects of the dictionary between brane configurations in del Pezzo
geometries and dibaryons in the dual superconformal quiver gauge
theories. The basis of fractional branes defining the quiver theory at
the singularity has a K-theoretic dual exceptional collection 
of bundles which can be used to read off the spectrum of dibaryons in the
weakly curved dual geometry. Our prescription identifies the R-charge
$R$ and all baryonic $\U(1)$ charges $Q_I$ with divisors in the del Pezzo
surface without any Weyl group ambiguity. As one application of the
correspondence, we identify the cubic anomaly $\tr R Q_I Q_J$ as an
intersection product for dibaryon charges in large-$N$ superconformal
gauge theories. Examples can be given for all del Pezzo surfaces using
three- and four-block exceptional collections. Markov-type equations
enforce consistency among anomaly equations for three-block
collections. 
}


\makepapertitle 

\body 

\version\versionno 

\tableofcontents

\newpage

\section{Introduction, summary of results, and outlook}

AdS/CFT dual pairs with $\caln=1$ supersymmetry have been much studied as natural
generalizations of the original duality \cite{jthroat,gkp,EW} between $\caln=4$ SYM and 
type IIB strings on ${\it AdS}_5\times {\S^5}$. After breaking of conformal invariance and 
supersymmetry, and combined with efforts at solving the strongly coupled worldsheets 
in small radius Anti-de Sitter space, they hold the promise of a holographic understanding 
of gauge theories describing the real world (see \cite{aharony} for a recent review).

A convenient way of engineering conformal $\caln=1$ dualities is to place a stack of
D$3$-branes at a conical Calabi-Yau threefold singularity \cite{LNV,KS,Kehag,Acharya,MoPle,KW}. 
To be specific, we will denote by $\bX$ the Calabi-Yau cone over the five-dimensional 
Sasaki-Einstein manifold $\Y$, which in the class of examples we consider is itself a 
circle bundle over a smooth K\"ahler-Einstein surface $\bV$ of positive curvature.%
\footnote{$\bV$ has dimension 2 and we add an extra leg for each additional direction.} 
The theory on D$3$-branes at the tip of $\bX$ is, at weak 't Hooft coupling, a quiver 
gauge theory, \ie, it has unitary gauge groups, bifundamental matter and a polynomial 
superpotential. At strong coupling, the theory is described by strings on ${\it AdS}_5
\times \Y$, with five-form flux on $\Y$. Breaking of conformal invariance can, in principle, 
be achieved on the gauge theory side by introducing fractional branes, and in the dual geometry 
by turning on additional fluxes leading to a warping of ${\it AdS}$ and to a deformation 
of $\Y$. In practice, of course, these deformations are rather difficult to study, and have 
been explicitly realized only in a very small set of examples, most prominently the conifold 
itself \cite{klst}. In this paper, we will consider only the conformal case, but we believe 
that our results will be very useful for future non-conformal deformations of the duality.

\subsection{Dibaryons}

One of the classical tests of AdS/CFT involves the matching of the spectrum and algebra of BPS 
states on the two sides of the duality. One of the entries in this dictionary relates
certain large-$N$ non-perturbative states in the gauge theory, \eg,
\begin{equation}
\epsilon_{i_1 i_2\ldots i_N} \epsilon^{j_1 j_2 \ldots j_N} X_{j_1}^{i_1}
X_{j_2}^{i_2}\cdots X_{j_N}^{i_N} \,,
\eqlabel{dibaryon}
\end{equation}
where $X_{j}^{i}$ is one of the bifundamental chiral matter fields in the quiver theory,
with D-branes wrapping various cycles in the dual geometry, for example, a D$3$-brane
wrapping a holomorphic curve in $\bV$ together with the $\U(1)$ fiber of $\Y$. Following
earlier literature \cite{Z3orb}, we will generically refer to these objects as dibaryons. 

The name dibaryon comes from the fact that the definition \eqref{dibaryon} 
involves the antisymmetrization over fundamental indices of two gauge groups. 
Recall that the baryon, a more fundamental object obtained by antisymmetrizing
over one gauge group, only exists when the theory is coupled to external charges \cite{EW2}.
In geometry, the baryon corresponds to a D$5$-brane wrapped on $\Y$ with fundamental strings
attached to it by the presence of the RR flux. In general, other states can be obtained 
by antisymmetrizing over more than two gauge groups, and should perhaps be called 
``polybaryons''. However, these objects can also be thought of as bound states of an 
equal number of baryons and anti-baryons (because the D$3$-brane is a bound state of 
a D$5$ and an anti-D$5$). In this sense, the denomination ``dibaryon'' is appropriate 
after all. 

These dibaryons were first studied in \cite{EW2} for the
geometry $\Y=\RP^5$ which is dual to a supersymmetric $SO(2N)$
gauge theory. This gauge theory has no bifundamental matter but does have
a field $\Phi_{ij}$ transforming in the adjoint of $SO(2N)$.
The Pfaffian $\mathrm{Pf}(\Phi)$ is a gauge invariant operator formed by
antisymmetrizing over $N/2$ copies of the $\Phi_{ij}$.  As the Pfaffian
consists of $N/2$ gluons, one expects it to have a mass of order $N$.
The string coupling $\lambda \sim 1/N$, and it seems logical to associate
the Pfaffian, as it has a mass of order $1/\lambda$ to a wrapped D-brane in the 
dual geometry. In particular, the Pfaffian is dual to a D3-brane wrapping a torsion 
cycle in the $\RP^5$.

Dibaryons in other simple geometries were studied soon after.  The authors of
\cite{Z3orb}, considered $\Y={\S^5}/\Z_3$ and the dual gauge theory ${\mathcal N}=1$ 
$\SU(N)^3$ Yang Mills with nine bifundamental matter fields $X_a$, $Y_b$, and $Z_c$,
three between each pair of $\SU(N)$ gauge groups. Dibaryons in $\Y=T^{1,1}$ were 
studied by \cite{KG}. The coset space $T^{1,1}$ is a level surface of the conifold. 
Here the gauge group is $\SU(N)^2$ and there are two types of bifundamentals $A_a$ 
and $B_b$ where $a,b=1,2$. More elaborate examples of dibaryons have been 
considered since then. For example, Beasley and Plesser \cite{Beasley1} give a 
detailed treatment of dibaryons for $\Y$ a $U(1)$ bundle over the third del Pezzo 
surface. The AdS/CFT dictionary for dibaryons provides one of the few
tests that AdS/CFT is a duality between string theory and gauge theory and not 
merely between supergravity and gauge theory.

Up until quite recently, work on dibaryons proceeded example by example.
In a recent advance, \cite{HM} describes how to calculate the dibaryon mass for
a general Sasaki-Einstein space $\Y$, given some information about the intersection 
of homology cycles in $H^2(\bV)$. In another intriguing recent paper \cite{IW2},
Intriligator and Wecht suggest that one should directly associate a divisor
in $H^2(\bV)$ to each bifundamental field $X^i_j$ at least in the case where 
all gauge groups are $SU(N)$. Such an association would greatly improve our 
understanding of the dibaryon spectrum, allowing direct comparison 
between antisymmetric products of the $X^i_j$ and the dual cycles wrapped by D3-branes. 
The genesis of the present paper was an attempt to make such an association
more precise and to understand where it could come from.

Thus, our primary goal is to describe a scheme that allows a complete and 
unambiguous identification between dibaryons in the quiver theory and branes 
wrapped on holomorphic curves in the dual geometry. Along the way, we will 
uncover an interesting structure that we believe can be applied to a variety
of other questions about this class of gauge/gravity dualities. In the
remainder of the introduction, we will give an overview of our results.
The details are split between the subsequent sections 2--7.

\subsection{Gauge theories at threefold singularities and exceptional collections}

Before the start, it is necessary to know the gauge theory. For orbifolds \cite{LNV,KS}
(where $\Y={\S^5}/\Gamma$), the field content and superpotential can be derived at weak 't 
Hooft coupling from perturbative string theory using the methods of Douglas and Moore 
\cite{DM}, and follows essentially from the representation theory of finite groups. 
For a certain limited class of other examples with non-spherical horizon, mainly toric 
del Pezzos, the gauge theories can be obtained by higgsing, or partial resolution, of 
orbifold singularities. These methods yield the gauge theories up to
ambiguities related to ``toric duality'' \cite{beasley0,fhh1}---which
can be understood in terms of Seiberg duality
\cite{Beasley1,fhh2}.

One can also take a slightly different point of view to study the
gauge theories at weak coupling and ask what happens when one resolves
the singularity and follows the D-branes to the corresponding large
volume Calabi-Yau manifold, an approach that has been developed in
recent years in particular in \cite{digo,DFR,dido,mayr,douglas}. This
method has the advantage that it can in principle be applied also to
non-toric geometries, such as the general del Pezzos. As we will
review below, the classical theory of exceptional collections of
bundles (or sheaves) appears naturally at large volume and is a
useful tool for organizing the gauge theories on branes at threefold
singularities. Exceptional collections also play an important role in 
the context of mirror symmetry, where they are mirror duals of certain
exceptional branes in Landau-Ginzburg models. A small list of
references is \cite{unify,zaslow,hiv,haiq}. In particular, exceptional
collections and their duals shed an interesting light on the
Cecotti-Vafa classification program of $\caln=2$ theories in two
dimensions \cite{ceva}, and it is conceivable that there is a relation
to our present work. The relevance of exceptional collections in
understanding the geometric structure of gauge theories through
gauge/gravity duality has been emphasized recently in
\cite{IW2,unify,wijn}. Here, we will see that the collections are also
useful for understanding the dibaryon spectra. 

We will now try to explain the relevant ideas
of\cite{digo,DFR,dido,mayr,douglas} in a popular example, the $\CZ$
orbifold. For a more pedagogical review, see \cite{trieste}. We will
review some material on exceptional collections in section \ref{exccoll}.

As is well-known, the $\CZ$ orbifold is continuously connected through variation of K\"ahler 
parameters to the large volume non-compact Calabi-Yau manifold $\calo_{\projective^2}
(-3)$ which is the total space of the canonical bundle over the complex projective 
plane $\projective^2$. 
We know from \cite{DM} that probe D3-branes on this Calabi-Yau experience an enhancement 
of gauge symmetry when the (closed string) K\"ahler modulus is tuned to the orbifold point. 
More precisely, the gauge group is enhanced from $\U(1)$ to $\U(1)^3$ and there 
are extra light fields $X_a,Y_b,Z_c$ appearing in bifundamental representations ($a,b,c=1,2,3$). 
The superpotential is of the well-known cubic form $\epsilon^{abc}X_a Y_b Z_c$. It is 
customary to view the D3-brane as the bound state of three elementary constituents, the 
fractional branes. In fact, it is claimed that all possible D-branes on $\CZ$ can be 
obtained as bound states of these three fractional branes \cite{DFR}. If this is true, 
and there are no antibranes involved, then all possible D-branes on the orbifold can 
be described as supersymmetric configurations in an $\caln=1$ field theory. As we 
move away from the orbifold point, supersymmetry is broken
spontaneously \cite{douglas}. In $\caln=2$ 
language (related to the previous language by three T-dualities in the spatial 
direction, whereby the D$3$-brane becomes a D$0$-brane in type IIA), the BPS central 
charges of the three fractional branes cease to be aligned. 

\begin{figure}
\begin{center}
\psfrag{SU(N)}{$\SU(N)$}
\psfrag{1}{$\!1$}
\psfrag{2}{$\!2$}
\psfrag{3}{$\!3$}
\psfrag{4}{$\!4$}
\psfrag{X}{$X_a$}
\psfrag{Y}{$Y_b$}
\psfrag{Z}{$Z_c$}
\epsfig{width=2.6in,file=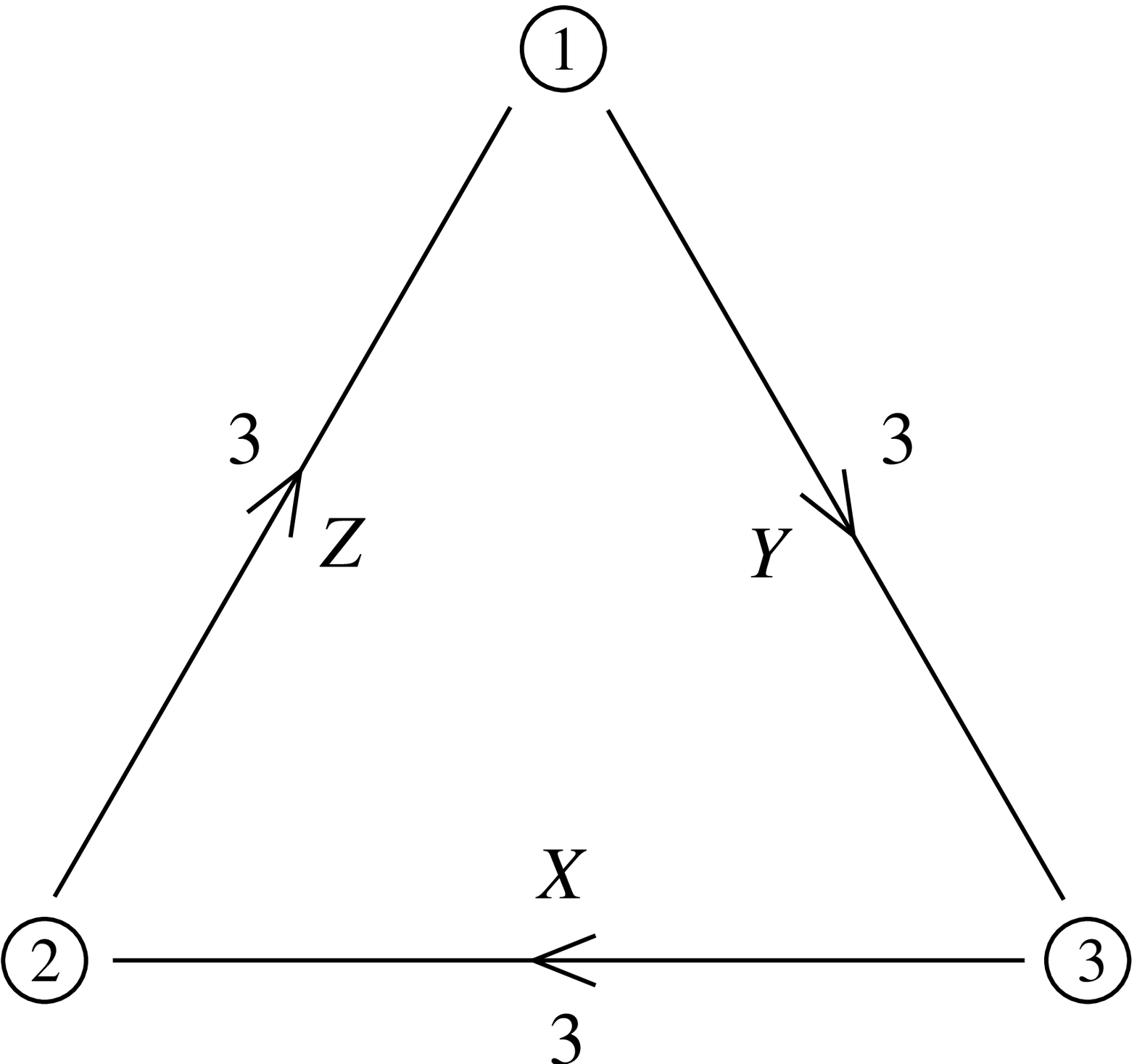}
\caption{Quiver for $\projective^2$.}
\label{figdp0}
\end{center}
\end{figure}

One of the important facts in this context is the decoupling statement \cite{douglas},
which is based on the topological twisting of the theory, and essentially states that 
holomorphic information in the worldvolume theory of the D-branes does in fact not 
depend on the K\"ahler parameters. In particular, the chiral spectrum and the holomorphic 
superpotential (F-terms) can be computed at large volume in classical geometry.
The K\"ahler parameters do, in the second step of finding supersymmetric
vacua, only affect the D-terms in the worldvolume theory.

Let us make these statements precise in the example. As first derived using 
mirror symmetry \cite{digo}, the three fractional branes $(e_1,e_2,e_3)$ of $\CZ$ are 
related at large volume to the collection of three bundles
\begin{equation}
\cale=(E_1,E_2,E_3)=(\calo(-1),\overline{T^*(1)},\calo) \,,
\eqlabel{exceptional}
\end{equation}
where $T^*(1)$ is the twisted cotangent bundle, and we have omitted introducing
a notation for the fact that we want to extend these bundles over $\projective^2$
to the Calabi-Yau.  (All ``bundles'' in the following discussion are sheaves with 
support on the compact space, and never extend in the non-compact directions.) 
In \eqref{exceptional}, we have used the overbar notation to denote that we have 
actually obtained the antibrane related to this bundle. 

As explained above, the chiral spectrum and the superpotential can now be easily 
computed from classical considerations. For instance, the massless fields 
between two branes defined by the bundles $E_1$ and $E_2$ are given 
by elements of the Dolbeault cohomology groups $H^{0,k}(E_1^*\otimes E_2)=\Ext^k
(E_1,E_2)$. Here, the star denotes the dual bundle and should not be confused with 
the overbar in \eqref{exceptional}. For the collection in
\eqref{exceptional} one finds that there are precisely three chiral
fields between each pair of bundles. The superpotential is computed by
multiplication of sections to be the familiar cubic superpotential,
and in this way, one has reproduced exactly the orbifold results. 

What these classical considerations cannot give is the D-terms in the gauge
theory. At large volume, the central charge
\begin{equation}
Z(E)= \int \EE^{-\omega} \ch(E) \sqrt{\Td(\bX)}
\eqlabel{central}
\end{equation}
of a brane depends only on the dimensionality of the corresponding submanifold 
(and the GSO projection), and is independent of the gauge bundle. However, at 
small volume, the formula \eqref{central} receives corrections from worldsheet 
instantons, which are responsible for the ``flow of gradings'' that maps the 
collection \eqref{exceptional} of two branes and one antibrane at large volume to 
the three fractional branes at the orbifold point with equal central
charge \cite{DFR,douglas}.

Let us note two remarkable facts about the gauge theory whose derivation from
large volume considerations we have just reviewed. First, there are no matter
fields in the adjoint representations of the gauge group, hence the gauge theory
does not have a Coulomb branch. Second, all matter fields between two fractional
branes are chiral. These two facts are precisely what makes the collection of 
bundles \eqref{exceptional} ``exceptional'' in the mathematical sense. In the 
language of bundles, adjoint matter fields would arise from non-trivial
morphisms from a bundle to itself; in other words, they would
correspond to deformations of the bundle. The chirality of the matter
means that, for fixed $i\neq j$, of all possible groups
$\Ext^k(E_i,E_j)$, at most one is non-zero, and in that case
$\Ext^k(E_j,E_i)$ vanishes for all $k$.  

Having reviewed the central ideas, we are now in a position to refer to the
recent work of Wijnholt \cite{wijn} for applications to the del Pezzo
surfaces $dP_n$ which will be the focus of our interest in
the subsequent sections. In particular, gauge theories have been
derived in \cite{wijn} for branes at the tip of cones over toric and
non-toric del Pezzos in a uniform manner, starting from exceptional
collections on $dP_n$. It is worthwhile to point out that one is
implicitly assuming that there actually is a point in the K\"ahler
moduli space at which the central charges of the exceptional
collections all align. The existence of such a point does not follow
from any results known to us. (Mirror symmetry is of no help here,
because the mirrors of the general del Pezzos are not known.) In
other words, it is not clear to us at what point, if any, in the
moduli space of $\bX=dP_n(K)$, the gauge theory actually lives. In what
follows, we will simply assume that there is such a point and that
semi-classically this point corresponds to a singular cone over the
del Pezzo. 

\subsection{Main results}

In this paper, we consider AdS/CFT dual pairs of theories which are obtained by 
choosing for $\bV$ a del Pezzo surface and for $\bX$ the complex cone over $\bV$. On the 
gauge theory side, we have the large-$N$ limit of the quiver gauge
theories, which one can derive from an exceptional collection on $\bV$
as we have briefly reviewed in the previous subsection. We 
would like an identification of the dibaryon operators in the field theory 
with branes in $\Y$. We claim that the spectrum of dibaryons follows in a natural 
way from an exceptional collection of bundles on $\bV$ ``dual'' to the 
collection defining the quiver theory. We  explain this ``duality'' in detail 
in section \ref{geodb}. In the example $\bV = \CP{2}$, the dual 
collection is 
\begin{equation}
\cale^\vee=(E_3^\vee,E_2^\vee,E_1^\vee)=(\calo,\calo(1),\calo(2)) \,.
\eqlabel{dual}
\end{equation}

Our proposal associates to each of the bifundamental fields in the quiver
a (fractional) homology class in $\bV$ which is the difference in 
the first Chern classes of the bundles on the corresponding nodes in the dual
collection.  Just as a collection of bifundamentals can antisymmetrize
to form a gauge invariant dibaryon, a collection of these fractional
homology classes can add up to an (integral) 
curve in $\bV$ which a D3-brane wraps together with the $\U(1)$ fiber
of the fibration $\Y\to\bV$. 

For $\CZ$, \eqref{dual}, our prescription simply associates with each of the bifundamental 
fields $X_a,Y_b,Z_c$ the hyperplane class in $\projective^2$, thus reproducing the
results of \cite{Z3orb}. For example, the fields $X_a$ run from node $3$ to node
$2$ in Fig.\ \ref{figdp0}, which in the dual collection \eqref{dual} correspond to
the bundles $\calo$ and $\calo(1)$, respectively. Thus, we associate with the dibaryon 
constructed out of $X_a$ the hyperplane divisor $H=c_1(\calo(1))-c_1(\calo)$.
The careful reader will suspect that our ordering and sign conventions must be 
quite complicated, and may also wonder about the field $Y$, which would get the 
divisor $-2H$ according to the rules as stated up to now. Indeed, a number of 
refinements will be necessary, and we will explain them in section \ref{geodb}.

This identification of baryon spectra is the central result of our paper. For 
completeness, let us also mention here a number of side results that are of 
interest in their own right.

Our $\caln=1$ superconformal gauge theories enjoy a number of global $\U(1)$
symmetries under which the dibaryons carry charge. On the AdS side, these 
symmetries are gauge symmetries arising from Kaluza-Klein compactification
on $\Y$. Our proposal includes a precise mapping of these $\U(1)$'s under
the duality.

The $\U(1)$ R-symmetry, for example, arises from the $\U(1)$ isometry of
$\Y$, and the R-charge of a dibaryon $B$ can be computed by intersecting the
corresponding curve $C$ with the canonical divisor in $\bV$, according to 
the formula \cite{HM,IW2}
\be
R(B) = \frac{2 N(-K)\cdot C}{K^2} \ ,
\eqlabel{r}
\ee
where $K$ is the canonical class of $\bV$.
For the $\CZ$ orbifold, we know that the anomalous dimensions of the
$X$, $Y$, and $Z$ fields vanish, so their R-charge is $2/3$. Antisymmetric
products of $N$ bifundamentals will have R-charge $2N/3$, and formula 
\eqref{r} reproduces this result ($K=-3H$ for $\projective^2$).

Other $\U(1)$ symmetries, so-called ``baryonic'' $\U(1)$'s, arise on the
AdS side from reducing the RR 4-form on three-cycles in $\Y$, which project to
divisors in $\bV$. Our prescription for identifying
dibaryons with curves allows a map from each baryonic charge $Q_I$ to a specific
divisor in $\bV$. One interesting point is that we are also able to identify
the gauge theory object that computes the intersection of divisors in $\bV$. 
This intersection product is simply the cubic anomaly $\tr R Q_I Q_J$. We will 
discuss these baryonic $\U(1)$'s and their intersection product in sections 
\ref{quiversec} and 4.

A different class of results in our paper---which are not all new---concerns 
algebraic conditions on anomalies in quiver gauge theories related to del 
Pezzos. We recall that, generally, vanishing of the chiral anomaly imposes 
certain restrictions on possible brane wrapping numbers. For the $n$-th del 
Pezzos $dP_n$ \cite{unify}, the restriction identifies an $n+1$
dimensional lattice of allowed brane wrappings, including both D$3$
and D$5$-branes. To find the D$3$-brane, one must identify which of
these configurations allows a vanishing NSVZ beta function, or  
equivalently, vanishing of the R-charge anomaly at each node in the quiver. The 
$n$ D$5$-branes wrapped on vanishing cycles break conformal invariance and 
do not allow an R-symmetry. There is a canonical one-to-one
correspondence between the $n$ D$5$-branes and the baryonic $\U(1)$'s
mentioned above  (see also \cite{IW2}).

Our strongest results are for the three-block exceptional collections.
Three-block collections give rise to quivers with three blocks of nodes.
Between two nodes in a block, there are no arrows. Karpov and Nogin
\cite{NK} have studied three-block collections over del Pezzo surfaces
and have shown that they are classified by certain diophantine
equations generalizing the Markov equation
\be
x^2 + y^2 + z^2 = 3xyz \,.
\eqlabel{markov1}
\ee
Eq.\ \eqref{markov1} is known \cite{unify} to classify all possible 
gauge theory quivers for the $\CZ$ orbifold. In particular, the $x$, $y$, 
and $z$ are the ranks of the three gauge groups. Karpov and Nogin \cite{NK} 
find a generalized Markov equation
\be
\alpha x^2 + \beta y^2 + \gamma z^2 = \sqrt{K^2 \alpha \beta \gamma} xyz
\eqlabel{markov2}
\ee
which classifies the three-block exceptional collections for del Pezzos. 
There are $\alpha$ nodes in the first block, $\beta$ nodes in the second, 
and $\gamma$ in the third and $x,y,z$ are the ranks as before. Through classifying 
three-block collections, this equation \eqref{markov2} also classifies
a large set of quivers for the corresponding gauge theories.

For three-block collections over del Pezzos, we show how the generalized Markov 
equation \eqref{markov2} arises as the consistency condition between the NSVZ 
beta function and R-charge of the superpotential.\footnote{This connection
was made for $\projective^2$ and \eqref{markov1} in \cite{wall}.} We
are also able to compute various cubic 't Hooft anomalies
explicitly. For example, AdS/CFT predicts that $\tr R^3 = 24 N^2 /K^2$.  We
are able to confirm this relation for all three-block collections. We
also check the maximizing-$a$ principle of 
\cite{IW}, \ie, that $\tr R^2 Q_I = 0$ for a superconformal field theory and 
that $\tr R Q_I Q_J$ has all negative eigenvalues, where $Q_I$
are the baryonic $\U(1)$'s mentioned above. Finally, we are able to show that 
$\tr Q_I Q_J Q_K  = 0$ for three-block collections.

To make the presentation less abstract, we will illustrate the working
of our machinery in a number of examples. We work out dibaryon spectra for four 
block quivers for $\CP{1}\times \CP{1}$, $\CP{2} $ blown up at a point, and 
$dP_2$.  We give an extensive treatment of quivers that arise from three block 
collections.  Finally, we work out all 240 smallest dibaryons for a three-block 
colllection for $dP_8$.\footnote{Martijn Wijnholt has also recently used the 
three-block exceptional collections of \cite{NK} for deriving quivers for $dP_7$ 
and $dP_8$.}

\subsection{Outlook}

We will see in this paper that exceptional collections, which appear naturally
in brane-geometric engineering of $\caln=1$ gauge theories,  are also useful 
for understanding the baryon spectra in AdS/CFT dual pairs arising from del Pezzo
geometries. The fact that the above identifications are highly non-trivial give 
us reason to expect that our results will shed light on a number of other related
questions, for example regarding non-conformal deformations of AdS/CFT.
We hope that the unambiguous mapping of gauge theory charges to cycles in 
the geometry will help us understand how to add fluxes
in the dual geometry to break conformal invariance in a controlled way.

We also feel that our results shed new light on the relation between
Seiberg duality and all its cousins (see
\cite{unify,fhh3,bedo,braun}). It was argued in \cite{unify} that
Seiberg duality is geometrically realized as mutations of the bundles
defining the gauge theory or as Picard-Lefshetz transformations in the
mirror geometry. Our results point to the importance of the
exceptional collection (\ref{dual}) dual to the bundles defining the
gauge theory. Indeed, Seiberg duality in the quiver gauge theory is
(in some cases) geometrically realized as a sequence of mutations on
this dual exceptional collection.

Finally, let us mention the extremely large class 
of largely unstudied examples where $\bX$ is a 
generalized conifold or more generally the cone over a log-del Pezzo.
In these cases, the associated surface $\bV$ is singular, but the
cone $\bX$ is smooth except at the tip, and very interesting quiver
gauge theories emerge, for example quivers identical to the
ADE Dynkin diagrams \cite{GNS}.  
Intersections of curves in $\bV$ can give rise to rational numbers, and it's
not clear how this rationality affects the charges of the dibaryons.
For the generalized conifolds, the quivers are non-chiral,
involving equal numbers of arrows in both directions between nodes.
It would be interesting to see how to generalize the notion of 
exceptional collections to these non-chiral examples.  Exceptional
collections as currently understood seem to allow only for chiral quiver theories.

\section{Exceptional collections and helices}
\label{exccoll}

As we have mentioned in the introduction, the notion of an exceptional collection of 
sheaves is a very natural mathematical concept in the context of D-branes. Among the recent 
physics literature, let us mention \cite{IW2,unify,hiv,wijn}. The standard mathematical 
reference is \cite{rudakov}. For the string theorist, an exceptional collection is simply a 
set of elementary ``rigid'' branes generating all BPS configurations of the theory by bound 
state formation. We have here summarized a few of the basic mathematical definitions 
to fix notation, but readers may want to skip the rest of this section.

Let $\bV$ be a complex Fano variety. A sheaf $E$ over $\bV$ is called exceptional if
$\Ext^0(E,E)=\complex$ and $\Ext^k(E,E)=0$ for $k>0$. An ordered collection $\cale=
(E_1, E_2, \ldots, E_m)$ of sheaves is called exceptional if each $E_i$ is
exceptional and if, moreover, for each pair $E_i,E_j$ with $i>j$, we have
$\Ext^k(E_i,E_j)=0$ for all $k$ and $\Ext^k(E_j,E_i)=0$ except possibly for a single 
$k$. 

For two sheaves $E$ and $F$ on $\bV$, we have the generalized Euler character
\begin{equation}
\chi(E, F) = \sum_i (-1)^i \dim \Ext^i(E,F) \,,
\eqlabel{Euler}
\end{equation}
which by the Hirzebruch-Riemann-Roch theorem can be rewritten as
\begin{equation}
\chi(E,F) = \int_\bV \ch(E^*) \ch(F) \Td(\bV) \,,
\eqlabel{HRR}
\end{equation}
where $\ch(E)$ is the Chern character of the sheaf $E$. The bilinear form
$\chi$ is non-degenerate on (the torsion-free part of) the K-theory
$K_0(\bV)$. 

It is easy to see from the definition that for an exceptional collection 
$\cale=(E_1,E_2,\ldots,E_m)$, the matrix $S$, with entries
\begin{equation}
S_{ij}=\chi(E_i,E_j) \,,
\eqlabel{S}
\end{equation}
is upper triangular with ones on the diagonal. Hence $S$ is invertible,
and it follows that the maximal length of an exceptional collection is
$m=\dim K_0(\bV)$. An exceptional collection which generates the derived
category of coherent sheaves on $\bV$ is called complete.

To be specific, we consider the del Pezzo surfaces, which are at the
center of interest in this paper. The $n$-th del Pezzo surface $dP_n$
has non-trivial Betti numbers $b_0=1$, $b_2=n+1$. Hence, $\dim K_0(dP_n)=n+3$,
and the maximal length of an exceptional collection is $n+3$. The Chern character
of a sheaf $\ch(E) = (r(E),c_1(E),\ch_2(E))$ is given by $n+3$ ``charges'': the rank
$r(E)$, the first Chern class $c_1(E)=\ch_1(E)\in H^2(dP_n,\zet)$, and the second
Chern character $\ch_2(E)$. In components, the Euler character reads
\begin{equation}
\begin{split}
\chi(E,F) &= r(E) r(F) + \frac{1}{2} \left( r(E) \deg(F) - r(F) \deg(E)
\right) \qquad\qquad\qquad \\
&\qquad\qquad\qquad + r(E) \ch_2(F) + r(F) \ch_2(E) - c_1(E) \cdot c_1(F) \,,
\end{split}
\eqlabel{chibasis}
\end{equation}
which can easily be derived from \eqref{HRR} using $\Td(dP_n)=1-\frac{K}2+H^2$,
where $K$ is the canonical class and $H$ is the hyperplane, with
$\int_{dP_n} H^2 = 1$.  Also, the degree $\deg(E) = (-K) \cdot c_1(E)$.

If $\cale$ is an exceptional collection, one obtains new exceptional collections
by so-called mutations. There are left and right mutations, so-called because of
the way they affect the ordering of the sheaves,
\begin{equation}
\begin{split}
L_i &: ( \ldots, E_{i-1}, E_i, E_{i+1}, \ldots ) \to
(\ldots, E_{i-1}, L_{E_i} E_{i+1}, E_i, \ldots ) \,, \\
R_i &: ( \ldots, E_{i-1}, E_i, E_{i+1}, \ldots ) \to
(\ldots, E_{i-1}, E_{i+1}, R_{E_{i+1}} E_i, \ldots ) \,.
\end{split}
\eqlabel{operate}
\end{equation}
Here, $L_{E_i}E_{i+1}$ and $R_{E_{i+1}} E_i$ are defined by short exact sequences,
whose precise form depends on which of the $\Ext^k(E_i,E_{i+1})$ is non-zero.
For exceptional collections over $dP_n$, only $\Ext^1$ and $\Hom$ are nontrivial
\cite{KO}.
For these mutations, 
if for an exceptional pair $(E,F)$, we have $\Ext^0(E,F)=\Hom(E,F)=
\complex^c$, where $c=\chi(E,F)$, we can consider the canonical mappings
\begin{align}
\Hom(E,F) \otimes E \to F  \eqlabel{leftcan} \ , \\
E\to \Hom(E,F)^*\otimes F \ ,
\eqlabel{rightcan}
\end{align}
and left and right mutations are defined as kernel or cokernel of these mappings.
So, if \eqref{leftcan} has a kernel, we have
\begin{equation}
0\to L_E F\to E{}^{\oplus c} \to F \to 0 \ .
\end{equation}
This particular mutation is called a division. The precise form of all the other 
mutations can be found, \eg, in \cite{rudakov}. 

We note that mutations operating on exceptional collections as in \eqref{operate}
satisfy the braid group relations
\begin{equation}
\begin{split}
L_i R_{i+1} = R_{i+1} L_i = 1 \ , \\
L_{i+1} L_i L_{i+1} = L_i L_{i+1} L_i \ .
\end{split}
\eqlabel{braid}
\end{equation}
Mutations are one of the important checks of the relation, via Mirror Symmetry,
between the theory of exceptional collections and Picard-Lefschetz theory for 
the singularities defining the associated mirror Landau-Ginzburg theories 
\cite{zaslow,hiv}. Moreover, we mention that in the context of constructing 
$\caln=1$ gauge theories from D-branes, a certain subclass of mutations is 
related to Seiberg duality \cite{unify,fhh3,oova}.

The most important aspect for us is the change in charges, which can be written 
as
\begin{equation}
\begin{split}
\ch(L_E F) &= \pm (\ch(F) - \chi(E, F) \ch(E) ) \,,\\
\ch(R_F E) &= \pm (\ch(E) - \chi(E,F) \ch(F) ) \,.
\end{split}
\eqlabel{charges}
\end{equation}
where the sign is chosen such that the rank of the mutated bundle is positive. 

In fact, in the context of D-branes in string theory, it is more natural to allow 
also negative charges (which is simply an antibrane, the formal inverse of a bundle 
with positive rank). One then defines an associated left and right mutation of
signed bundles $L^D$ and $R^D$, which at the level of charges leads to the selection
of the plus sign in \eqref{charges} in all cases. (The superscript $D$ refers to the 
derived category of coherent sheaves, to which the notion of exceptional collection 
is readily extended. We need these derived categories only 
implicitly here.) In those cases where
Seiberg duality is related to a sequence of mutations, one has to take into
account additional signs to keep the ranks of the gauge groups positive, see below.

Finally, let us consider a bi-infinite extension of an exceptional collection of 
sheaves $\cale=(E_1,E_2,\ldots,E_m)$, defined recursively by
\begin{equation}
\begin{split}
E_{i+m} &= R_{E_{i+m-1}} \cdots R_{E_{i+1}} E_i \\
E_{-i} &= L_{E_{1-i}} \cdots L_{E_{m-1-i}} E_{m-i} \qquad i\geq 0
\end{split}
\end{equation}
This infinite collection $\calh=(E_i)_{i\in\zet}$ is called a helix of period $m$ if 
\begin{equation}
E_i = E_{m+i} \otimes K \qquad\forall i\in\zet \,,
\end{equation}
where $K$ is the canonical bundle of $\bV$. Any subcollection of $\calh$ of the form 
$(E_{i+1},E_{i+2},\ldots,E_{i+m})$ is exceptional and is called a foundation for $\calh$. 
An interesting property of exceptional collections over Fano varieties (and del Pezzos 
in particular) is that an exceptional collection is complete (\ie, generates the derived 
category), if and only if it is the foundation of a helix.

\section{Quiver theories from exceptional collections}
\label{quiversec}

In this section we summarize the construction of the quiver theory
from an exceptional collection, discuss some properties of anomalies, and
explain the general construction of dibaryons in these field theories.

As we have outlined in the introduction (see also \cite{wijn}), we imagine starting from 
an exceptional collection $\cale=(E_1,E_2,\ldots,E_m)$ on the surface $\bV$ of interest. 
We extend this collection by zero to the total space of the canonical bundle $\bX=\bV(K)$ over 
$\bV$, which is Calabi-Yau, and we then follow this collection to the ``orbifold point'' in 
the moduli space of $\bX$. The collection $\cale$, in particular the grading, is chosen such 
that at the orbifold point, all $E_i$ correspond to mutually supersymmetric branes. We can 
then try to represent any brane $B$ on $\bX$ in a supersymmetric gauge theory, which is 
obtained as follows.

We start by drawing a quiver diagram, one node for each exceptional sheaf/frac\-tional 
brane in the collection, and an arrow between two nodes $i$ and $j$ whenever there is 
a non-zero $\Ext^k(E_i,E_j)$. Our convention for the direction of the arrow is that
it points from $i$ to $j$ if $\chi(E_i,E_j)$ is positive. We note that this sign depends 
not only on whether the arrow arises from an $\Ext^1$ or $\Hom$ at large volume, but also 
on the gradings that we were forced to chose. It is sometimes convenient to think of an 
arrow as an ordered pair $a=(ij)$, and to introduce the notation $t(a)=i$ for the
node at the tail of $a$ and $h(a)=j$ for the node at the head of $a$. We will sometimes
also use $w_a=\chi(E_i,E_j)\ge 0$ to denote the multiplicity or ``weight'' of the arrow 
$a=(ij)$.

Given the charge of any brane $B$ that we want to describe in gauge theory, we can
decompose its charge $\ch(B)$ in terms of the charges of the $E_i$,
\begin{equation}
\ch(B) = \sum_i d^i \,\ch(E_i) \,,
\eqlabel{decompose}
\end{equation}
where we assume that all $d^i$ are positive. We then associate in the gauge theory
a gauge group of rank $d^i$ to each node of the quiver, 
and chiral matter fields in bifundamental representations for each arrow (with chirality 
determined by the direction of the arrow). We will denote these fields by $X_a$ or by 
$X_{ij}$ for an arrow $a$ from $i$ to $j$.\footnote{We will generally suppress flavor indices
for arrows with multiplicity, because we will never break the non-abelian part of the 
flavor symmetry.}

The other piece of information we need is the superpotential. At large volume, the
superpotential encodes the relations between the morphisms from one exceptional bundle
to the other. Since the superpotential is independent of K\"ahler moduli, the relations 
between the chiral matter fields at the orbifold point will be the same as the ones 
obtained at large volume. For example, if $X_a\in\Hom(E_1,E_2)$, $Y_b\in\Hom(E_2,E_3)$,
appear with multiplicities, then the composition of maps $Y\circ X$
leads to maps from $E_1$ to $E_3$ which need not all be independent,
and the relations can be expressed by elements $Z\in\Hom(E_1,E_3)$. This
relation between three chiral  fields is encoded in a cubic term in
the superpotential. See \cite{unify,wijn} for additional  
details and examples. We note that in the quiver diagram, a superpotential term of order $r$ 
is associated to a closed loop of arrows $(a_1,a_2,\ldots,a_r)$ visiting the nodes $(i_1,i_2
\ldots,i_r)$, and by gauge invariance must be of the form $W_{a_1,a_2,\ldots,a_r}= \tr X_{a_1} 
X_{a_2}\cdots X_{a_r} = \tr X_{i_1i_2} X_{i_2i_3}\cdots X_{i_r i_1}$, with appropriately 
contracted color indices.

\subsection{Anomalies and non-conformal deformations}
\label{anomalies}

So far our description has been purely classical. We, however, are interested in studying
strongly coupled gauge theories, and we have to make sure that our theories make sense at
the quantum level. Cancellation of chiral gauge anomalies is equivalent to the condition 
that at each node, the number of matter fields in the fundamental and anti-fundamental be 
equal to each other. In terms of the matrix $S$, with entries $S_{ij}=\chi(E_i,E_j)$, this 
condition on the dimension vector $d=(d^1,d^2,\ldots,d^m)^t$ reads
\begin{equation}
\cali\,d=(S-S^t)\, d=0 \,.
\eqlabel{chiral}
\end{equation}
Given that  $\cali=S-S^t$ is the intersection form on the Calabi-Yau space $X$, the equation
\eqref{chiral} has the interpretation that from the closed string perspective, we are only 
allowed to wrap cycles that do not intersect any other compact cycle, because otherwise the 
flux sourced by the brane has nowhere to go \cite{unify}. For example, in the del Pezzo case, 
we have one compact 4-cycle, $n+1$ compact $2$-cycles, and one compact $0$-cycle in the 
geometry, which gives rise to $n+1$ linearly independent choices for wrapping branes. In 
other words, $\cali$ has rank $2$.

How about conformal invariance? As is well-known, conformal invariance of an $\caln=1$
supersymmetric gauge theory is tied to the existence of an anomaly-free $\U(1)$ R-symmetry.
In other words, to guarantee conformal invariance, we require that there be an assignment of 
R-symmetry charge $R(X_a)$ to each chiral multiplet $X_a$, such that at each node $i$, 
the NSVZ beta functions vanish,
\begin{equation}
2 d^i + \sum_a w_a (R(X_a)-1) \bigl(\delta^i_{t(a)} d^{h(a)} + \delta^i_{h(a)}d^{t(a)}\bigr)=0
\qquad\text{for every node $i$} \,,
\eqlabel{NSVZ} 
\end{equation}
and that moreover, each term in the superpotential have R-charge $2$
\begin{equation}
R(X_{i_1i_2}) + R(X_{i_2 i_3}) + \cdots + R(X_{i_r i_1}) = 2 \qquad\text{\parbox[t]{6cm}{for 
every loop $i_1,i_2,\ldots,i_r$ in the superpotential.}}
\eqlabel{supo}
\end{equation}
In \eqref{NSVZ}, $R(X_{a})$ is the R-charge of the bottom component. Eqs.\ \eqref{NSVZ} 
and \eqref{supo} are a system of inhomogeneous linear equations on the R-charges of 
the $X_{ij}$. Existence of a solution puts certain constraints on the possible 
numbers $d^i$. For the del Pezzos, we expect from string theory that out of the 
$n+1$ possible brane wrappings, only the regular D$3$-brane at the orbifold will
respect conformal symmetry, and all fractional branes will break it. In other
words, we expect that out of the $n+1$ solutions of \eqref{chiral}, exactly one of
them will allow for an R-symmetry.

In general, there will also be other anomaly-free $\U(1)$ flavor symmetries.
We can distinguish ``mesonic'' $\U(1)$ symmetries that do not assign the same charge 
to each of the several fields corresponding to one given arrow, in other words, that do 
not commute with the non-abelian part of the flavor symmetries \cite{Beasley1}, from 
``baryonic'' $\U(1)$ symmetries that do commute with the non-abelian part of the flavor
symmetries.\footnote{The fact that the R-symmetry behaves like the baryonic $\U(1)$'s
follows from the maximizing-$a$ principle of \cite{IW}.} Baryonic $\U(1)$ charges, which 
we denote by $Q_I$, have to satisfy the homogeneous versions of \eqref{NSVZ} and 
\eqref{supo},
\begin{equation}
\begin{split}
\sum_a w_a Q_I(X_a) \bigl(\delta^i_{t(a)} d^{h(a)} + \delta^i_{h(a)}d^{t(a)}\bigr)
&=0 \qquad \text{for every node, and}\\
Q_I(X_{i_1 i_2}) + Q_I(X_{i_2 i_3}) +\cdots +Q_I(X_{i_r i_1}) &=0 \qquad\text{for every loop.}
\end{split}
\eqlabel{homogeneous}
\end{equation}

We claim that the number of these baryonic charges is in general exactly one less than the 
number of solutions to eq.\ \eqref{chiral}, and that moreover, for given choice of dimension
vector $d_*$ satisfying \eqref{chiral}, the baryonic $\U(1)$'s are in one-to-one 
correspondence with the other solutions of \eqref{chiral}. To show this, let us define a 
complete basis of charges $Q_a$ that commute with the non-abelian part of the flavor symmetries
by $Q_a(X_b)=\delta_{ab}$. Any solution of \eqref{homogeneous} can be written as a combination 
of the $Q_a$, $Q_I=\sum_a q_I^a Q_a$. From the condition that the superpotential have charge 
$0$, and assuming that there are enough independent terms in the superpotential, we infer 
that $Q_I$ must give charge $0$ to any field consisting of an incoming arrow at some node 
followed by an arrow coming out of that node. In other words, $Q_I$ must be a linear combination 
of the charges $Q_i$ that assign charge $+1$ to each arrow going into the node $i$ and $-1$ to 
each outgoing arrow, $Q_I = \sum_i q_I^i Q_i$ with
\begin{equation}
Q_i = \sum_a ( \delta_{i, h(a)} - \delta_{i, t(a)}) Q_a \,.
\eqlabel{Qi}
\end{equation}
Using this definition, it is easy to see that the first condition in \eqref{homogeneous}
becomes
\begin{equation}
\sum_j \cali_{ij} q_I^j d^j_* = 0 \qquad\text{for every node $i$.}
\end{equation}
Thus, if we denote by $d_I=(d_I^i)$ the solutions of \eqref{chiral}, we can write the 
solutions to \eqref{homogeneous} as
\begin{equation}
q_I^i = \frac{d_I^i}{d^i_*}  \,,
\eqlabel{baryonic}
\end{equation}
assuming that for our selected dimension vector $d^i_*$ is non-zero for all $i$. We
also note that for $d_I=d_*$, the corresponding $\U(1)$ is trivial, because it
assigns charge $0$ to all fields.  This justifies the claim
we made at the beginning of
this paragraph.

The condition that $d^i$ must be non-zero for all $i$ for the baryonic charges to exist
has a tempting interpretation based on the expectation about the RG behavior of our
theories. Let us assume that the conformal theory (stemming from D$3$-branes at a point) 
has all $d^i$ positive. As we have just seen, the baryonic $\U(1)$'s
in this theory are in one-to-one correspondence with the possible
non-conformal deformations of the theory \cite{IW2}. Following
\cite{klst}, we can make a non-conformal deformation by adding
fractional branes, and the deformation will be controllable if the
number of fractional branes is much smaller than the number of regular
D$3$-branes. Controllable means 
in particular that all $d^i$ remain positive and the baryonic charges
still exist, although they now slightly differ from the conformal case. The theory 
will then start flowing with the scale, and it is natural to expect a cascade similar 
to the one of Klebanov and Strassler \cite{klst}. In particular, the number of D$3$-branes will 
decrease along the cascade, until one eventually reaches a confining theory, where
some of the $d^i$ become zero, and the baryonic $\U(1)$'s might disappear. On the
gravity side, the RG flow and cascade correspond to a warping of the geometry, with 
sizes of various cycles in $\Y$ depending on the radial direction. Ultimately, confinement 
in the gauge theory is expected to correspond to a deformation of the singularity,
with certain cycles smoothly shrinking to zero size in the geometry. In other words, 
the condition that the $d^i$ must be positive for the baryonic $\U(1)$'s to exist is 
related to the existence of cycles in the level surfaces of the dual geometry, and 
some of the baryonic $\U(1)$'s will disappear in the IR when the gauge theory confines.
In fact, it is shown in \cite{unify} that deformation of the geometry is possible
precisely because some of the $d^i$ vanish for certain non-conformal deformations of 
the theory involving a small number of regular D$3$-branes.

Let us return to the conformal case. If for some choice of $d$, there is an assignement
of R-charge that satisfies \eqref{NSVZ} and \eqref{supo}, then this solution will
not be unique because of the non-trivial solutions of \eqref{homogeneous}. To fix this 
ambiguity, it is natural to first look for additional discrete symmetries of the quiver 
diagram that fix the R-charges of certain matter fields to be equal. However, in general 
these constraints do not suffice. Luckily, Intriligator and Wecht 
have recently shown \cite{IW} that there is a completely general method to fix the ambiguity. 
The prescription is that the exact superconformal R-charge is, among all solutions of \eqref{NSVZ} 
and \eqref{supo}, distinguished by the fact that it maximizes $a$, which is one of the 
central charges of the conformal algebra of the SCFT \cite{anselmi1,anselmi2},
\begin{eqnarray}
a = \frac{3}{32} \bigl(3\tr R^3 - \tr R \bigr) &=& \frac{3N^2}{32}
\Bigl(\sum_{a} w_a d^{h(a)} d^{t(a)}\bigl[3 (R(X_{a})-1)^3 - 
(R(X_{a})-1)\bigr] \nonumber \\
&& \; \; \; \; \; \; +2 \sum_i (d^i)^2 \Bigr)\,,
\label{a}
\end{eqnarray}
where the second term is the contribution from the gauginos. As we have mentioned, in some 
cases the exact R-symmetry is uniquely fixed by symmetries, but this is not true for an 
arbitrary quiver in which not all discrete symmetries are manifest. In the subsequent
sections, we have made frequent use of the maximizing-$a$ principle to determine the R-charge.

\subsection{Dibaryons in quiver gauge theories}
\label{diba}

{}From now on, we will consider the conformal case only. This means that at each node $i$, we 
put a gauge group of rank $N d^i$, where the vector $(d^1,d^2,\ldots, d^{n+3})^t$ is the (unique) 
null vector of $\cali$ that admits an R-symmetry, and $N$ is large. Our goal is to provide 
certain kinematical tests of $\caln=1$ AdS/CFT by matching BPS observables on the two sides 
of the duality. At weak coupling, the observables in question are gauge invariant combinations 
of chiral operators. One possibility are traces. These are perturbative also in the large $N$ 
expansion, and correspond on the closed string side to gravitons or other supergravity fields 
\cite{EW}. On the other hand, determinant-like states, such as \eqref{dibaryon}, are 
non-perturbative at large $N$ (because their mass scales as one over the coupling constant)
and correspond on the closed string side to branes. 

For a convenient description of general dibaryon operators, let us for the moment just keep 
the vector space structure of the quiver, as we would do for the purposes of solving the F-terms
or in the mathematical setting. In other words, we associate to each node $i$ a vector space $V_i$
of dimension $N d_i$, and to each chiral field a linear map $X_{ij}:V_i\to V_j$. These linear
maps furnish a representation of the path algebra $\cala$ of the quiver diagram. We recall that the
algebra structure means that products of paths are non-zero whenever arrows line
up head to tail, and the associated map is simply a composition of maps between vector spaces. 
We can also consider sums of maps in the usual way.

Let us then consider an arbitrary element of the path algebra $A\in\cala$
\begin{equation}
A : V_t \mapsto V_h \,,
\eqlabel{def}
\end{equation}
where 
\begin{equation}
V_t=\oplus_i t^i V_i \qquad\qquad V_h =\oplus_i h^i V_i 
\end{equation}
are the vector spaces at the ``tail'' and ``head'' of $A$, respectively. Note that $A$ need 
not consist of only one path, and that some of the vector spaces $V_i$ can appear with non-trivial 
multiplicity $t^i$, $h^i$ in the tail and head of $A$. Moreover, we note that a non-trivial 
multiplicity $w_a>1$ of some arrow $a$ will give rise to a collection of different maps in 
general (unless there are relations from the superpotential).

\begin{figure}[t]
\begin{center}
\psfrag{V1}{$V_1$}
\psfrag{V2}{$V_2$}
\psfrag{V3}{$V_3$}
\psfrag{V4}{$V_4$}
\psfrag{V5}{$V_5$}
\psfrag{X}{$X$}
\psfrag{Y}{$Y$}
\psfrag{Yp}{$Y'$}
\psfrag{Z}{$Z$}
(a) 
\hskip -.7cm
\raisebox{1cm}{\epsfig{width=5cm,file=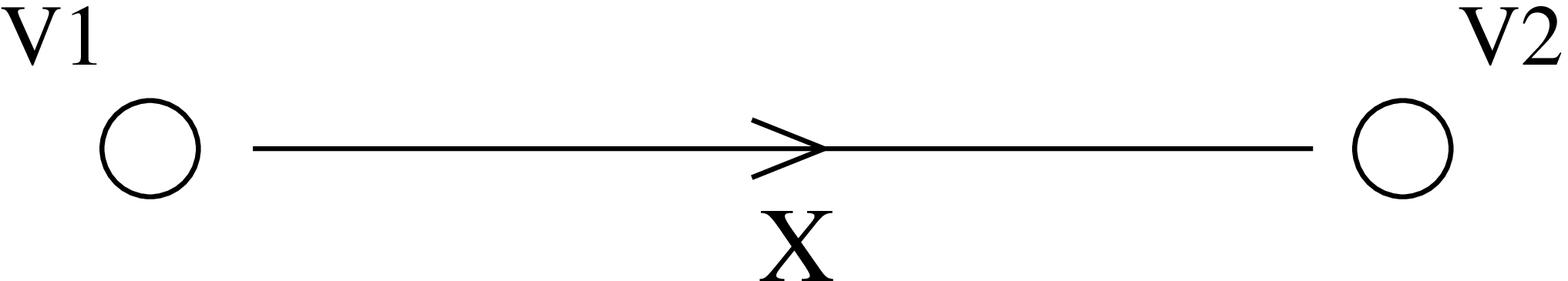}}
\hfil
(b) 
\epsfig{width=5cm,file=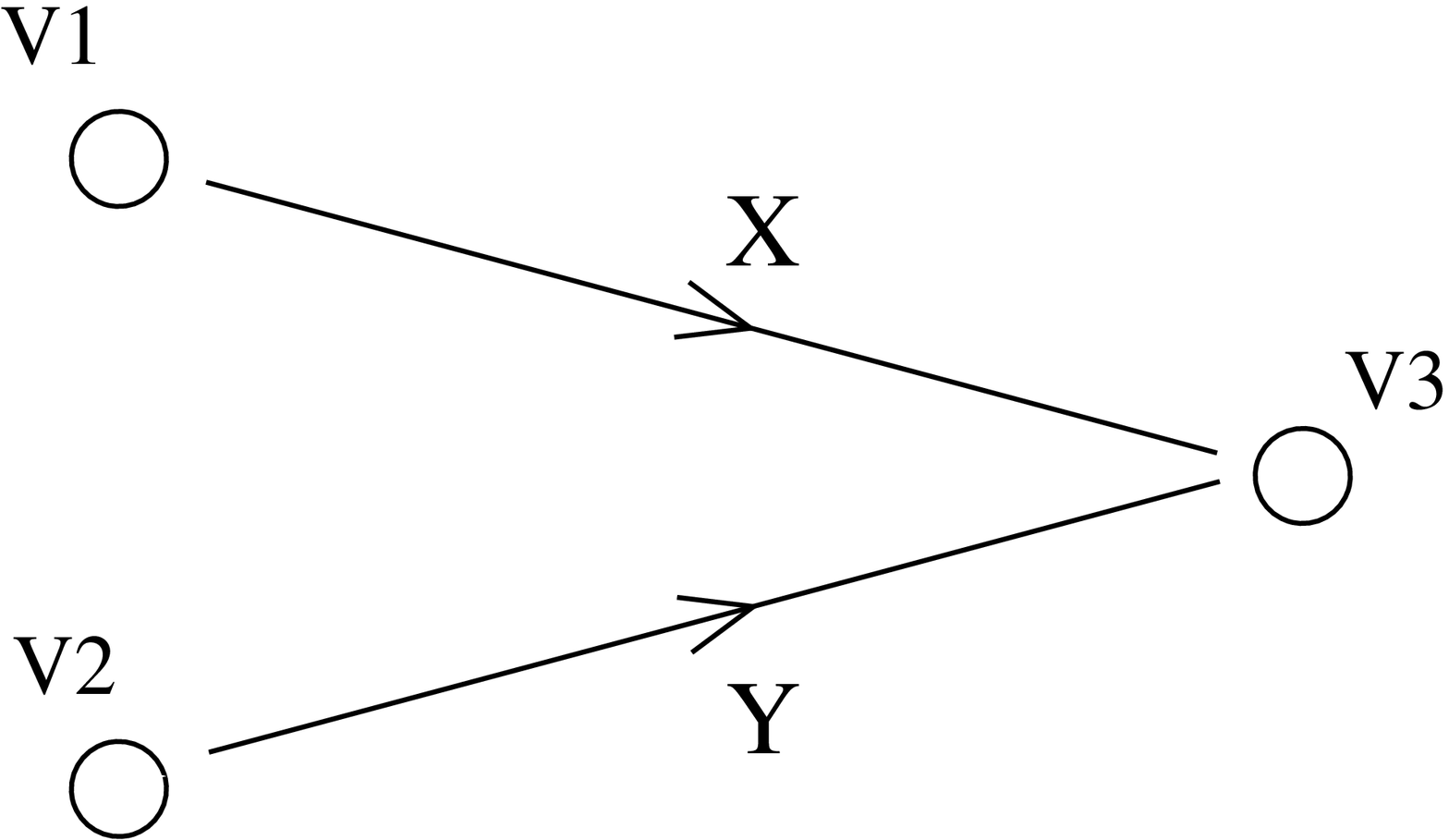}
\vskip .5cm
(c) 
\hskip -.7cm
\epsfig{width=5cm,file=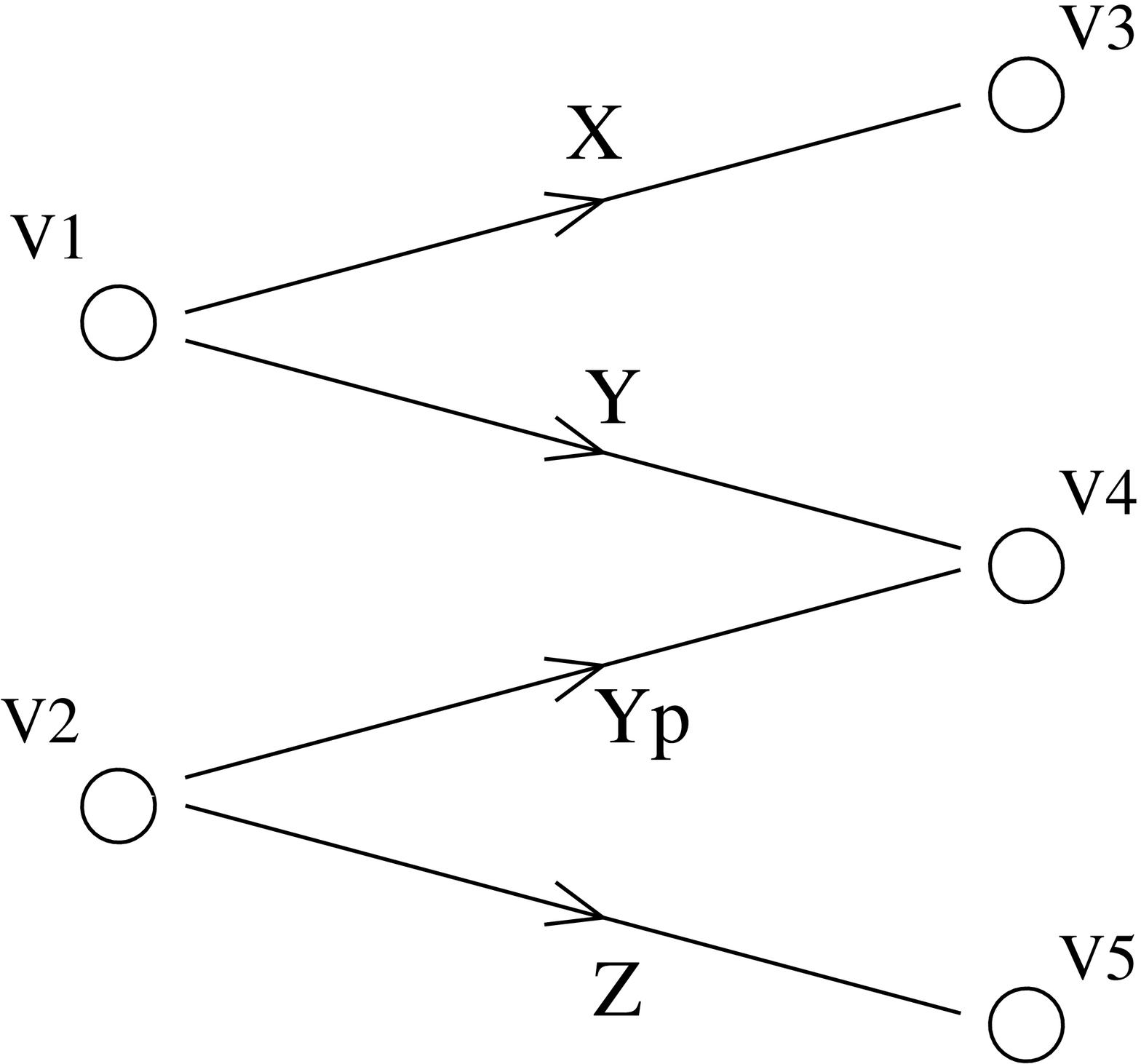}
\hfil
(d) 
\hskip -1cm
\raisebox{.7cm}{\epsfig{width=10cm,file=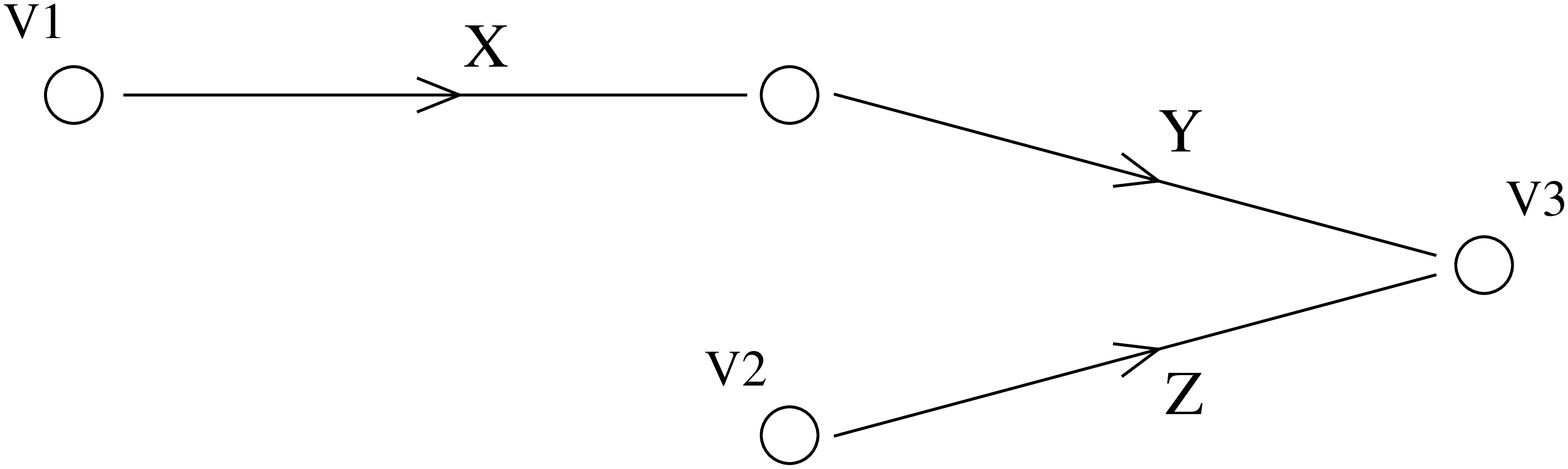}}
\caption{Examples of subquivers that can be used to construct dibaryon operators
in the field theory.}
\label{di}
\end{center}
\end{figure}

A useful invariant, which we call the ``rank'' of $A$, is the
difference in dimension of the vector spaces at tail and head,
\begin{equation}
r(A) = \dim(V_h)-\dim(V_t) = \sum_i N d^i (h^i-t^i) \,.
\eqlabel{rank}
\end{equation}
In gauge theory, $r(A)$ counts the number of uncontracted fundamental minus antifundamental indices in
each gauge group. In order to form a gauge invariant operator out of $A$, we have to contract these
open indices with external charges and take an antisymmetric combination over all the remaining ones.
In the special case that $r(A)=0$, we can simply define this antisymmetrized product, the dibaryon, as 
the determinant of the linear map associated with $A$, \footnote{One might worry that the determinant of 
a map between different vector spaces is not well-defined. However, we are here interested in gauge 
theory, so we will need a hermitian metric on each $V_i$. We can then define the determinant of the 
map as the determinant of its matrix 
representation with respect to some unitary basis. This definition 
is independent of the choice of unitary basis. We thank Andrei
Mikhailov for raising this question.}
\begin{equation}
B(A) = \det A \,.
\eqlabel{trivial}
\end{equation}
We depict a few examples of dibaryons constructed along these lines in Fig.\ \ref{di}.
In case (a), we have 
\begin{equation}
\dim V_1=\dim V_2=N 
\qquad  B=\det X\,,
\end{equation}
in case (b), we have
\begin{equation}
\begin{split}
\dim V_1 &=N \\
\dim V_2 &=N 
\end{split}
\qquad
\begin{split}
\dim V_3&=2N
\end{split}
\qquad\qquad 
B=\det \begin{pmatrix} X & Y \end{pmatrix}\,,
\end{equation}
for (c), the dibaryon satisfies
\begin{equation}
\begin{split}
\dim V_1 &=3N\\ \dim V_2&=3N
\end{split}
\qquad
\begin{split}
\dim V_3&=2N\\
\dim V_4&=2N\\
\dim V_5&=2N
\end{split}
\qquad\qquad
B= \det\begin{pmatrix} X & 0 \\ Y & Y' \\ 0 & Z \end{pmatrix} \,,
\end{equation}
and, finally, for (d),
\begin{equation}
\begin{split}
\dim V_1&=N \\
\dim V_2&=N
\end{split}
\qquad\quad
\begin{split}
\dim V_3=2N 
\end{split}
\qquad\qquad B= \det \begin{pmatrix} YX & Z \end{pmatrix}\,.
\end{equation}
One advantage of this condensed notation in terms of determinants---instead of the 
explicit notation in terms of $\epsilon$-symbols---is that it is quite easy to see
which paths give factorized dibaryons, or which paths do not actually 
yield any dibaryon at all because the determinant is zero. A zero determinant can
occur, for example, if the multiplicities of the vector spaces is larger than the number 
of arrows between them, so that the same arrow must appear several times. Moreover,
it is also quite easy to formulate deformations of dibaryons. For instance, if between
$V_t$ and $V_h$ we have two maps $A_1$ and $A_2$, we can consider the determinant of
the general linear combination
\begin{equation}
B(\lambda_1,\lambda_2)= \det (\lambda_1 A_1+\lambda_2 A_2) \,.
\eqlabel{moduli}
\end{equation}
Note that we must require $|\lambda_1|^2+|\lambda_2|^2=1$ for the dibaryon to be
properly normalized and that we identify modulo the phase $(\lambda_1,\lambda_2)
\equiv \EE^{\ii\phi}(\lambda_1,\lambda_2)$. In other words, the moduli of these
dibaryons is the projective space $\projective^1$. The existence of such a moduli
space is also a typical situation on the geometry side. In quantizing this moduli space,
as explained in \cite{Z3orb}, the coupling to the RR flux leads to the identification 
of the ground state Hilbert space with the space of sections of $\calo_{\projective^1}(N)$, 
which has dimension $N+1$. In  field theory, these ground states correspond to 
the homogeneous terms in the expansion of $B(\lambda_1,\lambda_2)$ in powers of 
$(\lambda_1,\lambda_2)$, and there are also exactly $N+1$ of them.

\section{The Geometry of dibaryons}
\label{geodb}

In AdS/CFT,
dibaryons correspond to D$3$-branes wrapped on supersymmetric 3-cycles in 
$\Y$ which project to holomorphic curves in $\bV$.
On the gauge theory side, 
because they are constructed as antisymmetrized products of bifundamental chiral
fields, the dibaryons of our quiver theories carry charges $Q_I(B)$ under the
baryonic $\U(1)$ symmetries discussed in subsection \ref{anomalies}. Moreover,
they also carry a definite R-charge $R(B)$. Geometrically, the $\U(1)$
charges are naturally identified with the homology class of the
corresponding curves. 

We review briefly a construction from \cite{HM, Beasley2, Mikhailov}
in order to better understand the stability of these D3-brane wrappings.  
We start with a Euclidean D3-brane wrapping a holomorphic, 
four-dimensional cycle in the Calabi-Yau cone $\bX$. The total space is 
$\R{4} \times \bX$. Now we move to a $H_5 \times \Y$ geometry by adding a 
large amount of five-form flux.  The space $\Y$ is five dimensional and is 
Sasaki-Einstein.  The space $H_5$ is Euclidean $AdS_5$ and is highly
symmetric. In particular, we can choose the old radial direction
of $\bX$ to Wick rotate.  After Wick rotation, $H_5$ will turn into
$AdS_5$.  The lowest energy dibaryons are typically time independent.
Tracing back through the above construction, we see that time
independent wrappings corresponded initially to D3-brane
wrappings that were independent of the radial coordinate.
Moreover, the wrappings were holomorphic and the radius
is paired holomorphically with the $U(1)$ coordinate
in the Sasaki-Einstein space $\Y$.  Thus these wrappings
are also independent of the $U(1)$ angle.  In other words,
time independent dibaryons wrap holomorphic curves $C \subset \bV$ 
along with the $U(1)$ fiber above every point in $C$.

In the geometric identification of the 
global $U(1)$ symmetries,
the simplest charge to understand is the R-charge. 
Roughly speaking one expects the mass of the D$3$-brane
to be proportional to the tension times the volume of this wrapped 
three-cycle in $\Y$. Moreover, for objects in AdS/CFT correspondence 
with mass of order $N$, the mass and conformal dimension are
equal up to corrections of order $1/N$.  Finally, dibaryons are 
chiral primary operators so we expect that their conformal dimension 
$\Delta = 3R/2$, where $R$ is the R-charge. On the other hand,
because $\bV$ is Einstein, the volume of the curve $C$ is 
proportional to its intersection with the canonical class $K_\bV$
of $\bV$, \ie, to its degree 
$\deg(C)= - K_\bV \cdot C$. Taken together, the following exact 
result of \cite{HM, IW2} should not be surprising,
\begin{equation}
R(B) = \frac{2N}{K_\bV^2} \deg(C) \,.
\label{dibaryonr}
\end{equation}
The constant of proportionality comes from carefully
considering the D3-brane tension and the metric on ${\it AdS}_5 
\times \Y$.  The fact that the R-charge, and not the mass shows
up on the left hand side of the equation was discussed by 
\cite{BHK}, and comes from the existence of fermionic zero 
modes in $AdS_5$ for these dibaryons.

Beyond the R-charge, one would like also to identify all other
baryonic $\U(1)$ charges $Q_I$ unambiguously with geometric
quantities. As discussed in section 3, these other $U(1)$'s are
naturally associated to the remaining generators of $H^2(\bV)$, but it
is not yet clear which $U(1)$'s correspond to which elements of the
homology class. We fix this ambiguity in section \ref{secid} by
identifying curves $C$ with antisymmetric products of bifundamental
fields. We also discuss a number of consequences and then provide
extensive tests for $\bV$ a del Pezzo surface. First, however, we
discuss the intersection form on the curves $C$ in greater detail.

\subsection{Charges and intersection form}
\label{intersec}

One interesting property of curves in surfaces is that they intersect in points,
and it is natural to ask what could be the analog of the intersection product on
the gauge theory side. In search for such a quantity, let us rewrite the 
maximizing-$a$ principle \eqref{a} as the condition
\begin{equation}
\tr R^2 Q_I = 0 \qquad \text{for all $I$,}
\eqlabel{null}
\end{equation}
which ensures that we are at a critical point of $a$, together with
\def\inter{{\mathfrak K}}
\begin{equation}
\inter<0,\quad\text{where $\inter$ is the matrix $\inter_{IJ}=\tr R Q_I Q_J$\,,}
\eqlabel{negative}
\end{equation}
which ensures that we have a maximum. In fact, the maximizing-$a$ principle was derived
from precisely these two conditions in \cite{IW}. The matrix $\inter_{IJ}$ is symmetric because 
the $Q_I$ commute with $R$, and it is tempting to interpret $\inter$ as an intersection form. 
We will see that this identification is indeed correct, at least if $\bV$ is a del Pezzo 
surface. More precisely, we want to view $\tr R Q_1 Q_2$ as the intersection of the two charges 
$Q_1$ and $Q_2$, which can be the generators of the $\U(1)$ R-symmetry or of any of the baryonic 
$\U(1)$'s. Then \eqref{null} says that the R-charge is orthogonal to the baryonic $\U(1)$'s while
\eqref{negative} is the statement that the orthogonal complement of the R-charge has a 
definite signature. On the del Pezzo side, the corresponding statement is that the orthogonal 
complement of the canonical divisor $K$ is negative definite, and is in fact isomorphic to the 
root lattice of the exceptional simply laced Lie algebras, see, \eg, \cite{mysterious}. 
Via AdS/CFT, the canonical divisor corresponds to the R-charge and the 
orthogonal complement to the baryonic $\U(1)$'s.

To make this explicit, let us denote by $C(B)$ the holomorphic curve that corresponds to the
dibaryon $B$. The projection orthogonal to the canonical divisor $K$ is given by
$C^\perp= C - K (K\cdot C)/K^2$. The formula that we will compare between
gauge theory and geometry is then explicitly
\begin{equation}
C^\perp(B_1)\cdot C^\perp(B_2) = \frac 12 Q_I(B_1) \inter^{IJ} Q_J(B_2)\,,
\eqlabel{compare}
\end{equation}
where the LHS is the geometric intersection product in the surface $\bV$, while the
RHS is a gauge theory expression in terms of the $\U(1)$ charges of two dibaryons
and the inverse $\inter^{IJ}$ of the matrix 
$\inter_{IJ}$.\footnote{%
This identification of $\tr RQ_I Q_J$ as an intersection form is reminiscent of a similar 
formula on the string worldsheet. In the context of D-branes on Calabi-Yau spaces (in fact, 
also in flat space), it is well-known that the intersection of two D-branes $D_1$ 
and $D_2$ can be computed from the Witten index in the Ramond sector of open strings 
stretched between $D_1$ and $D_2$\,,
\begin{equation*}
D_1\cdot D_2 = \tr_{\calh^{\rm R}_{D_1 D_2}} (-1)^F \,,
\tag{$*$}
\end{equation*}
see, \eg, \cite{hiv,dofi}. Upon modular transformation to the closed string
sector, the expression ($*$) becomes the overlap of the corresponding
boundary states
\begin{equation*}
D_1\cdot D_2 = \langle D_1|\EE^{-\pi\ii J_0}\EE^{-2\pi\tau H_{\rm cl})} 
|D_2\rangle_{\rm RR}
\tag{$**$}
\end{equation*}
in the Ramond-Ramond sector with insertion of the R-charge, with only ground
states contributing in the limit of $\tau\to\infty$. Our formula \eqref{compare} can 
be viewed as the holographic version of eq.\ ($**$). It would be interesting 
to see whether there is also a gauge theory expression for ($*$).}

\subsection{Dual exceptional collections}

We recall that the gauge theory of interest was obtained starting
from an exceptional collection of bundles $\cale=(E_1,E_2,\ldots,E_{m})$, 
where $m=n+3$ is $\dim K_0(\bV)$ of the $n$-th del Pezzo surface $\bV=dP_n$.
To each $E_i$ corresponds a node in the quiver and to each $\Ext^k(E_i,E_j)$
a chiral field $X_{ij}$. A dibaryon $B$ is constructed by antisymmetrization 
from a certain combination of the $X_{ij}$. Our basic claim is that one can 
read off the class of the curve corresponding to $B$ using not the exceptional
collection $\cale$, but a certain dual exceptional collection, $\cale^\vee$,
which we define presently.  It is important to point out that the original 
collection $\cale$ lives on the large volume Calabi-Yau (and, by continuity, 
on the singular cone $\bX$), while the dual collection $\cale^\vee$ lives on
the large-$N$ dual geometry $\Y\to \bV$.

For any exceptional collection $\cale = (E_1, E_2, \ldots E_m)$, we define a
dual collection $\cale^\vee$ as the result of a certain braiding operation,
\begin{equation}
\begin{split}
{\mathcal E}^\vee 
&= (E_m^\vee,E_{m-1}^\vee,\ldots,E_1^\vee) \\
&= (L^D_{E_1} \cdots L^D_{E_{m-1}} E_m, 
L^D_{E_1}  \cdots L^D_{E_{m-2}} E_{m-1}, \ldots,
L^D_{E_1} E_2, E_1) \,.
\end{split}
\eqlabel{evee}
\end{equation}
The collection $\cale^\vee$ is exceptional in the order presented, and is dual to 
$\cale$ in the sense of the Euler form, \ie, $\chi(E_i, E_j^\vee) = \delta_{ij}$. 
This property of the collection \eqref{evee} is easily checked based on the 
transformation of the charges under braiding, eq.\ \eqref{charges}, linearity of 
$\chi$, and the upper triangularity of the matrix $S_{ij}=\chi(E_i,E_j)$.

We note that dual exceptional collections play an important role also in the 
weak coupling limit in the context of the McKay correspondence \cite{dido,mayr}, 
and in mirror symmetry \cite{hiv}. The role they play in AdS/CFT is new, but has 
been hinted at in \cite{IW2,mrd}. 

One interesting property of the dual collection is that the ranks of the dual
bundles are proportional to the ranks $Nd_i$ of the gauge groups in the original 
quiver. This follows from the fact that the charge of the D$3$ brane, which is the
class of a point in $\bV$, decomposes as\footnote{From now on, $d_i\equiv d^i$.}
\be
\sum_i d_i \ch(E_i) = \ch(\mO_p) \,,
\eqlabel{ch}
\ee
and multiplying with $\chi(\cdot,E_i^\vee)$ yields the equality 
$d_i=r(E_i^\vee)$. In particular, since all $d_i>0$,
all $E_i^\vee$ are actual bundles (and not anti-bundles), as befits a
large volume description. Moreover, multiplying \eqref{ch} with 
$\chi(\mO_p,\cdot)$ yields
\be
\sum_i r(E_i^\vee) r(E_i) =
\sum_i \chi(\mO_p, E_i^\vee)
\chi(E_i, \mO_p) = \chi(\mO_p, \mO_p) = 0 \ .
\ee
We can also investigate the $\chi$-dual of a sequence of mutations on
$\cale=(E_1,E_2,\ldots,E_n)$. Using \eqref{braid}, it is easy to see that
\begin{equation}
(L_i\cale)^\vee = L_{n-i} \cale^\vee \,,
\end{equation}
which implies that for $j>i$
\be
\left( L_{j-1} L_{j-2}  \cdots L_{i+1} L_i {\mathcal E} \right)^\vee =
L_{n-j+1} \cdots L_{n-i-1} L_{n-i} ({\mathcal E}^\vee) \ ,
\eqlabel{mutated}
\ee
This relation is interesting because in some cases the sequence
of mutations on the LHS of \eqref{mutated} is equivalent to Seiberg duality in the
gauge theory.  The RHS of \eqref{mutated} then gives a straightforward way to 
determine the ranks of the gauge groups and charges of the fields after Seiberg
duality.

\subsection{Dibaryon charges from exceptional collections}
\label{secid}

To identify the charges of the dibaryons in geometry, it will suffice to 
specify the classes of the generating fields $X_{ij}$. We note that the 
classes that we associate with the $X_{ij}$ will in general be fractional,
and that only the combinations of $X_{ij}$ that upon antisymmetrization
give non-trivial dibaryons on the field theory side will correspond to 
integral classes with actual curves as representatives.

Consider an arrow $a=(ij)$ connecting two nodes in the quiver,
with corresponding bifundamental field $X_{ij}$. The sheaf that is
dual to the node at the tail is $E_{t(a)}^\vee=E_i^\vee$, and the 
sheaf that is dual to the head is $E_{h(a)}^\vee=E_j^\vee$. We distinguish 
two cases. Either $\chi(E_i^\vee,E_j^\vee)\neq 0 $ or $\chi(E_i^\vee, 
E_j^\vee)=0$. In the first case, the fractional divisor is 
\be
D_a=D_{ij}=\frac{c_1(E_j^\vee)}{r(E_j^\vee)} - 
\frac{c_1(E_i^\vee)}{r(E_i^\vee)} \ .
\eqlabel{divisor}
\ee
In the second case, we add $-K$ to the divisor: $D_a \to D_a-K$.
In all cases we have studied so far, the resulting
divisor has positive degree, $D_a \cdot (-K) > 0$. This is reassuring,
since we know that the R-charge in field theory is related to the
degree by $R(X_a) = 2 \deg(D_a) /K^2$, and we want the R-charges of all
chiral fields to be positive. 

To see how the formula \eqref{divisor} associates integral classes
with dibaryons, recall that on the field theory side
we need to take enough and appropriate kinds of bifundamental
matter fields to antisymmetrize completely over the color
indices. As explained in subsection \ref{diba}, the
condition is that the dimension of the vector spaces at head and tail
be equal. Since $d_i=r(E_i^\vee)$, the condition that there is no
uncontracted fundamental index is that the differences of ranks in the dual
collection be zero, \ie, that there is no D$5$-brane charge. In the
simplest case where we antisymmetrize over only one bifundamental matter 
field, the dibaryon contains $N \lcm (d_i,d_j)$ copies of the field
(which are distinguished by their flavor index). Recalling from 
the previous subsection that $d_i=r(E_i^\vee)$, we see that the
factor $\lcm (d_i,d_j)$ is of just the right form to place the curve 
$C$ corresponding to the dibaryon in the integer homology
\be
C = \frac{r(E_i^\vee) c_1(E_j^\vee) - r(E_j^\vee) c_1(E_i^\vee)}
{\gcd(r(E_i^\vee), r(E_j^\vee))}\ .
\ee

This example captures the spirit of the process of
constructing dibaryons, but is an oversimplification. Even
though the coefficients in $C$ may be integers, the
divisor may still not be good for wrapping branes.
Recall that we may compute the genus of a curve from
the adjunction formula $2g-2 = C\cdot(C+K)$. Many times, an 
arbitrary $C$, even if integral, would have $g<0$!
In practice, we have often found that when $g<0$,
there is some corresponding gauge theory obstruction
which makes the corresponding antisymmetrization vanish,
for example because $w_a$ is too small. We do not, however, know a
general proof why the dibaryon should vanish whenever there is no
corresponding curve to wrap. 

Using \eqref{divisor}, we can give a geometric and simple 
formula for the R-charge of the bifundamental matter fields, 
obtained by intersecting $D_{ij}$ with $-K$,
\be
R(X_{ij}) = \frac{2}{K^2 r(E_i^\vee) r(E_j^\vee)}  \times
\left\{
\begin{array}{cl}
\chi(E_i^\vee,E_j^\vee) & \; \; \; \mathrm{if} \; \; 
\chi(E_i^\vee, E_j^\vee) \neq 0 \\
\chi(E_i^\vee\otimes K,E_j^\vee) & \; \; \; \mathrm{otherwise.}
\end{array} \right.
\eqlabel{rcharge}
\ee
where $E_i^\vee$ and $E_j^\vee$ live in the dual collection.
The constraint that the R-charge of a chiral field be positive
becomes the condition that  
$\chi(E_i^\vee, E_j^\vee) \geq 0$ and  
$\chi(E_j^\vee, E_i^\vee) <  r(E_j^\vee) r(E_i^\vee) K^2$. 

This R-charge formula \eqref{rcharge} has a remarkable property under
reversal in arrow direction. If we denote by $X_{ji}$ the antichiral
field conjugate to $X_{ij}$, then \eqref{rcharge} yields $R(X_{ij}) =
2-R(X_{ji})$, as one can easily check using \eqref{chibasis}.
Recall that the fermions in the chiral multiplet have R-charge
$R_f = R-1$.  Thus, under this switch in arrow direction,
$R_f \to -R_f$, as one would have expected.

For the reader interested where this identification between
bifundamentals and fractional divisors comes from, 
it builds on work of \cite{IW2} and
was arrived at by studying a large number of examples, some of which
we will describe in the following sections.  However, having intuited the 
relationship between the $X_{ij}$ and the lattice of divisors, we
can now go back and make sure these formulae have the right properties.

First, we can check that loops in the quiver, which can correspond to
terms in the superpotential, can have R-charge two.  The R-charge of such
a loop is proportional to the degree of the sum of the fractional divisors 
$D_a$ over all arrows $a$ in that loop,
\be
R_{\mathrm{loop}} = \frac{2}{K^2} \sum_{a\in \mathrm{loop}} d(D_a) \ .
\ee
Moreover, from the definition of the fractional divisors,
the sum over the $D_a$ must be an integer multiple of $c_1(\bV) = -K$
(all the other classes cancel in the sum). Therefore,
\be
R_{\mathrm{loop}} = 2 n \ .
\ee 
Finally, from the structure of the exceptional collection, $n \geq 1$.
We conclude that the loop can appear in the superpotential only
if $n=1$.

Second, we can test that (\ref{rcharge}) satisfies the NSVZ beta functions.  
The NSVZ beta functions \eqref{NSVZ} can be written for each node $i$ as
\be
\beta_i = 
2d_i^2 + \sum_j R_f(X_{ij}) d_i d_j (\chi(E_i, E_j) - \chi(E_j, E_i)) \ .
\ee
After some manipulation, this expression reduces to
\be
\beta_i = 
\frac{2}{K^2} \sum_k (\chi(E_i^\vee, E_k^\vee)-\chi(E_k^\vee, E_i^\vee))
(\chi(E_i, E_k) - \chi(E_k, E_i)) + 2r(E_i^\vee) r(E_i) \ .
\ee
To show that the $\beta_i$ do indeed vanish, we will show something stronger,
namely that the matrix ${\it NSVZ}_{ij}$ vanishes identically, where
\be
{\it NSVZ}_{ij} = 
\frac{2}{K^2} \sum_k (\chi(E_i^\vee, E_k^\vee)-\chi(E_k^\vee, E_i^\vee))
(\chi(E_j, E_k) - \chi(E_k, E_j)) + 2r(E_i^\vee) r(E_j) \ .
\eqlabel{nij}
\ee

The proof can be described easily and falls into two pieces.  First, one shows
that under the basis transformation $F_j = B_{ij} E_i$, 
\be
{\it NSVZ} \to B^{-1} \cdot ({\it NSVZ}) \cdot B \ .
\ee
Second, one shows that ${\it NSVZ}$ vanishes in a particular basis.  
A convenient choice is the basis in which the sheaves are expressed
in terms of their rank, first Chern class, and second Chern
character, i.e. the basis in which $\chi$ is written as in
(\ref{chibasis}).  In this basis $2r(E_i^\vee) r(E_j)$ is
a matrix which is zero everywhere except for a two in the
lower left hand corner.  The first term in \eqref{nij} is similarly 
a matrix which is zero everywhere except for a $-2$ in the lower left 
hand corner.

In the same way that $R(X_{ij})$ was obtained by intersecting the
fractional divisor $D_{ij}$ with $K$,
the $Q_I(X_{ij})$ are obtained by intersecting the 
$D_{ij}$ with the generators of the lattice orthogonal to $K$. It would
be ideal to finish this section with a demonstration that
this choice of $R_{ij}$ maximizes the conformal anomaly
$a$ \cite{IW} over the space of $Q_I$
for a general quiver and corresponding exceptional
collection, but we have as yet only
been able to prove it for three block exceptional collections
and on a case by case basis.

We now turn to the examples.

\section{Simple examples of exceptional collections}

Having tackled the surface $\CP{2}$ in the introduction, we consider
dibaryons arising from holomorphic curves in 
the slightly more complicated surfaces $\CP{1} \times \CP{1}$ and
$\CP{2}$ blown up at one or two points.

\subsection{The First del Pezzo}

There are in fact two del Pezzos with $K^2=8$.  One is
$\CP{1}\times \CP{1}$ and the other is $\CP{2}$ blown
up at a point.  The sheaves on both surfaces are
easy to describe.  The second Betti number of both
surfaces $b_2=2$ and so we need two weights to
describe the lattice of divisors.

We begin with $\CP{1} \times \CP{1}$.  A divisor
can be written $D=mf + ng$ where $f\cdot f=0$,
$g \cdot g = 0$, $f \cdot g = 1$, and $m$, $n\in \Z$.
The canonical divisor $K= -2f - 2g$.
Exceptional collections on $\CP{1}\times \CP{1}$
are well known.  For example, \cite{unify} gives the
collection 
$(\mO(-f-g), \overline{\mO(-f)}, \overline{\mO(-2g)}, \mO(-g))$.
In this basis, $S_{ij}$ is indeed upper triangular
\be
S = 
\left(
\begin{array}{rrrr}
1 & -2 & 0 & 2 \\
0  & 1 & -2 & 0 \\
0 & 0 & 1 & -2 \\
0 & 0 & 0 & 1 
\end{array} 
\right) \ .
\ee
The dual collection is ${\mathcal E}^\vee =
(\mO(-2f-3g), \mO(-2f-2g), \mO(-f-2g), \mO(-f-g))$ and the
corresponding $(S^{-1})_{ij}$ is
\be
S^{-1} = 
\left(
\begin{array}{cccc}
1 & 2 & 4 & 6 \\
0  & 1 & 2 & 4 \\
0 & 0 & 1 & 2 \\
0 & 0 & 0 & 1 
\end{array}
\right) \ .
\ee
{}From the ranks of the dual bundles, we infer that the ranks of the
gauge groups in the quiver are all $SU(N)$.  

We can also read off the R-charges of the generators of
the algebra.
{}From (\ref{rcharge}), we see that the R-charges of the generators
are all $1/2$. This charge agrees with the possible 
dibaryon R-charges which are integer multiples of $N/2$.

We can also compute the ``fractional'' divisors, which are in
fact not fractional in this example, for each 
bifundamental field.  These divisors are nothing but
$f$ and $g$.  Antisymmetrizing over the bifundamental
fields corresponds to constructing the holomorphic
curve $C = af + bg$ where $a$ and $b$ are non-negative
integers.

\begin{figure}
\psfrag{SU(N)}{$\SU(N)$}
\psfrag{1}{$1$}
\psfrag{2}{$2$}
\psfrag{3}{$3$}
\psfrag{4}{$4$}
(a)
\hskip 0.1cm
\epsfig{width=2.6in,file=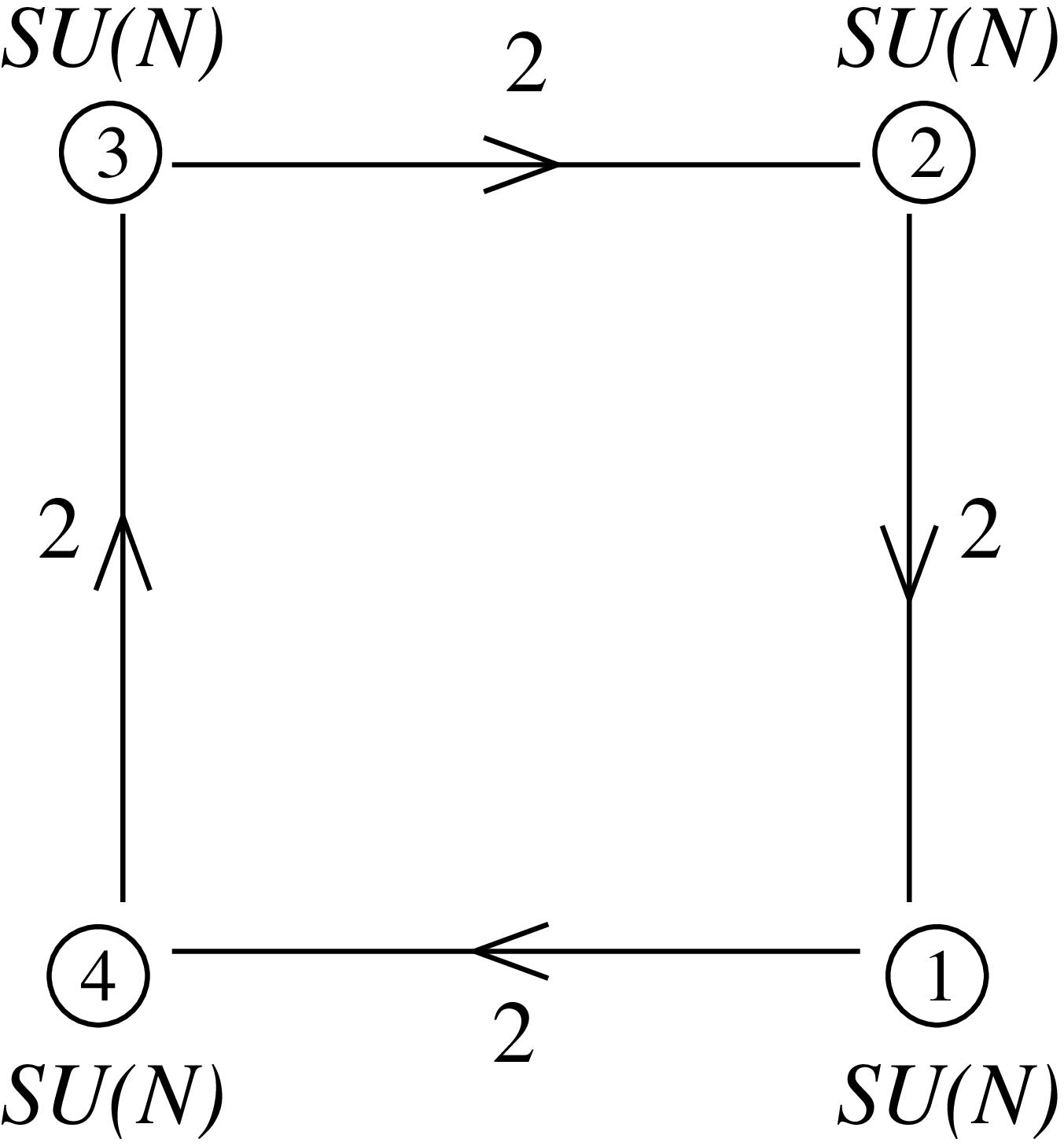}
\; \; \; 
(b)
\epsfig{width=2.45in,file=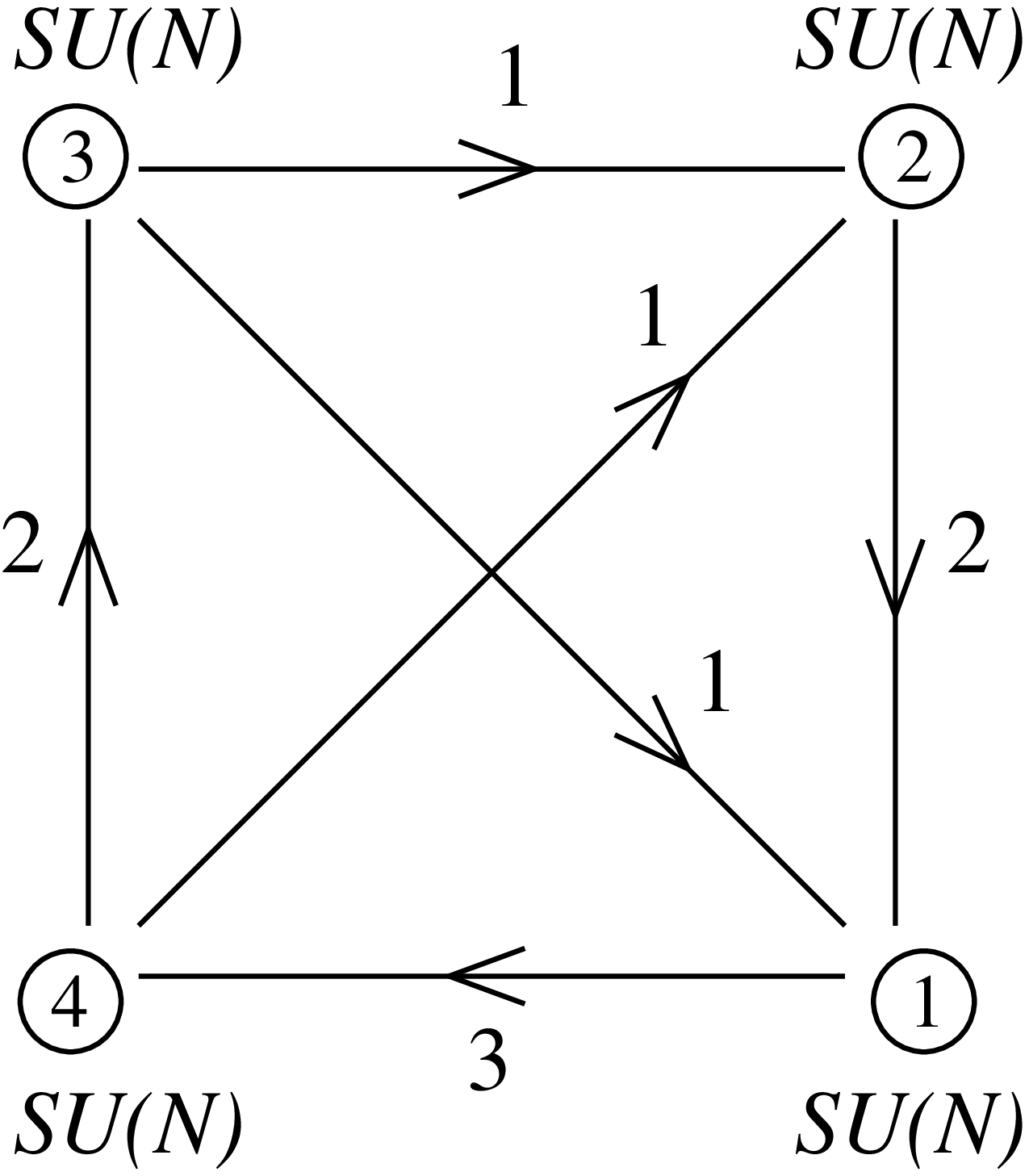}
\caption{Quivers for a) $\CP{1} \times \CP{1}$ and
b) $\CP{2}$ blown up at a point.}
\label{figdp1}
\end{figure}

The next example is $\CP{2}$ blown up at a point.
We denote by $H$ the hyperplane of $\CP{2}$ and by $E$
the exceptional divisor. The most general divisor can
then be written $D= aH + bE$ where $a$, $b \in \Z$ and
$H\cdot H=1$, $E\cdot E=-1$, and $H\cdot E=0$.
Exceptional collections are well known for this example as
well.  We take ${\mathcal E} =
(\mO, \overline{\mO(H-E)}, \overline{\mO(E)}, \mO(H) )$ from \cite{unify}.
In this basis 
\be
S = 
\left(
\begin{array}{rrrr}
1 & -2 & -1 & 3 \\
0  & 1 & -1 & -1 \\
0 & 0 & 1 & -2 \\
0 & 0 & 0 & 1 
\end{array} 
\right) \ .
\ee
The dual collection is
${\mathcal E}^\vee = (\mO(-2H+E), \mO(-H), \mO(-H+E), \mO)$
and 
\be
S^{-1} = 
\left(
\begin{array}{cccc}
1 & 2 & 3 & 5 \\
0  & 1 & 1 & 3 \\
0 & 0 & 1 & 2 \\
0 & 0 & 0 & 1 
\end{array} 
\right) \ .
\ee
The ranks of the gauge groups are again all $SU(N)$ for this quiver.
Moreover, we can read off the R-charges and fractional
divisors of the generators
\be
\renewcommand{\arraystretch}{1.2}
\begin{array}{ccc}
 & \mathrm{D} &{\mathrm{R}} \\
X_{32} & E & \frac{1}{4} \\
X_{21} & H-E & \frac{1}{2} \\
X_{43} & H-E & \frac{1}{2} \\
X_{42} & H & \frac{3}{4} \\
X_{31} & H & \frac{3}{4} \\
X_{14} & H & \frac{3}{4}  \ .
\end{array}
\ee
These same R-charges can be calculated
by maximizing the conformal anomaly $a$ 
\cite{IW}.  
Moreover, this table 
agrees with the table (4.13) of \cite{IW2} where 
the authors arrived at the $D$ in a rather different
way.

\subsection{The Second del Pezzo}

A divisor for the second del Pezzo ($\CP{2}$ blown up by two exceptional 
divisors $E_1$ and $E_2$) may be written $D = aH + b_1 E_1 + b_2 E_2$ where 
$a$ and $b_i$ are integers and $H\cdot H=1$, $E_i \cdot E_j = -\delta_{ij}$,
and $H \cdot E_i = 0$.

\begin{figure}[t]
\begin{center}
\psfrag{SU(N)}{$\SU(N)$}
\psfrag{SU(2N)}{$\SU(2N)$}
\psfrag{1}{$1$}
\psfrag{2}{$2$}
\psfrag{3}{$3$}
\psfrag{4}{$4$}
\psfrag{5}{$5$}
\epsfig{width=3.5in,file=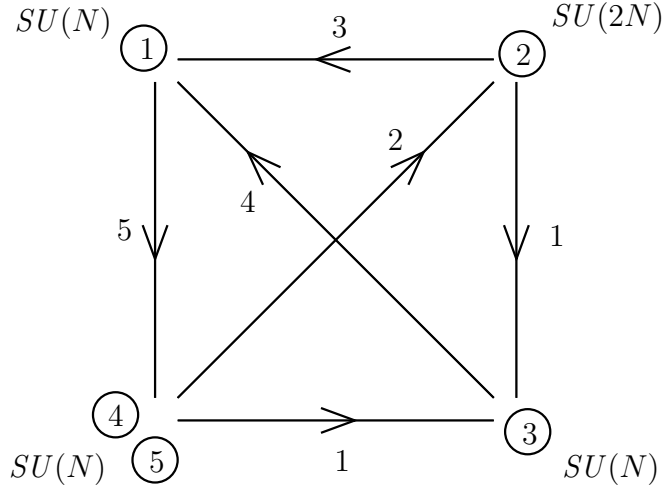}
\end{center}
\caption{Quiver for the second del Pezzo.}
\label{figdp2}
\end{figure}

Exceptional collections are well known for the second del Pezzo.
One such collection is
${\mathcal E} = (\mO, \overline{\mO(H)}, 
\overline{\mO(2H-E_1-E_2)}, \mO(2H-E_1),
\mO(2H-E_2))$.  In this basis
\be
S = 
\left(
\begin{array}{rrrrr}
1 & -3 & -4 & 5 & 5 \\
0  & 1 & 1 & -2 & -2 \\
0 & 0 & 1 & -1 & -1 \\
0 & 0 & 0 & 1 & 0 \\
0 & 0 & 0 & 0 & 1 
\end{array} 
\right) \ .
\ee
We show the quiver corresponding to $\cale$ in Fig.\ \ref{figdp2}.
The inverse matrix is
\be
S^{-1} = 
\left(
\begin{array}{rrrrr}
1 & 3 & 1 & 2 & 2 \\
0  & 1 & -1 & 1 & 1 \\
0 & 0 & 1 & 1 & 1 \\
0 & 0 & 0 & 1 & 0 \\
0 & 0 & 0 & 0 & 1 
\end{array} 
\right) \ .
\ee
This example is the first we have come across so far that requires
the use of higher rank bundles.  In particular, the dual collection is
${\mathcal E}^\vee = (\mO(-H+E_1), \mO(-H+E_2), 
\mO(-H+E_1 + E_2), F, \mO)$ where the charges of $F$
are $\ch(F) = (2, -H, -1/2)$.  Because the rank of $F$ is two,
there will now be one gauge group in the quiver, corresponding
to $\overline{\mO(H)} \in {\mathcal E}$, 
with gauge group $SU(2N)$ instead of
$SU(N)$. 

Using the formulae for the R-charges
and the fractional divisors, we find that
\be
\renewcommand{\arraystretch}{1.2}
\begin{array}{ccc}
 & D & {\mathrm{R}} \\
X_{42} & \frac{1}{2} H - E_2 &  \frac{1}{7} \\
X_{52} &  \frac{1}{2} H - E_1 & \frac{1}{7} \\
X_{31} & H-E_1-E_2 & \frac{2}{7} \\
X_{43} & E_1 & \frac{2}{7} \\
X_{53} & E_2 & \frac{2}{7} \\
X_{21} & \frac{1}{2} H & \frac{3}{7} \\
X_{14} & 2H - E_1 & \frac{10}{7} \\
X_{15} & 2H - E_2 & \frac{10}{7} \\
X_{23} & \frac{5}{2} H & \frac{15}{7}. 
\end{array}
\eqlabel{dp2charges}
\ee
These same R-charges can be obtained by
maximizing the conformal anomaly $a$ with some
weak assumptions about the form of the superpotential.

It is fun to see how these fractional divisors combine to give
a dibaryon.  From gauge theory, there is a dibaryon
that can be constructed from an antisymmetric product
of $N$ $X_{42}$ fields and $N$ $X_{52}$ fields.  
{}From the geometric perspective, we see that
the fractional divisors for these fields combine to give
the curve $H-E_1-E_2$, a degree one curve
corresponding to one of the smallest wrapped
D3-branes in the dibaryon spectrum.

We can also check our formula \eqref{compare} for the intersection of 
the dibaryons. In the $Q_a$ basis for the charges, where we take
the ordering of arrows to be as in (\ref{dp2charges}), the two baryonic 
$\U(1)$ charges satisfying \eqref{homogeneous} are given by
$Q_1=(-3,-1,-4,2,4,1,2,0,5)$ and $Q_2=(1,-1,0,1,-1,0,-1,1,0)$.
Using the R-charges in \eqref{dp2charges}, one then finds
\begin{equation}
\inter = \tr R Q_I Q_J=\begin{pmatrix} -32 & 4 \\ 4 & -4 \end{pmatrix}\,.
\eqlabel{dp2inter}
\end{equation}
The charges $Q_1$ and $Q_2$ correspond geometrically to the intersection 
with the divisors orthogonal to the canonical class $-3H+E_1+E_2$. It is easy
to check that
\begin{equation}
\begin{split}
& 2H- 4 E_2 -2 E_1\,, \\
& E_2-E_1
\end{split}
\end{equation}
indeed have the same mutual intersection products as in \eqref{dp2inter} 
(up to a factor of $2$), and that the intersection with the fractional
divisors in \eqref{dp2charges} are the same as the charges of the 
corresponding chiral fields.

Let us note that in principle, one could have tried to derive the identifications
in \eqref{dp2charges} just by looking at the charges of the fields and their 
intersection product in field theory (assuming one knew about the
significance of $\tr RQ_I Q_J$). However, the identification of this data
with the geometry is ambiguous because of the discrete symmetries shared
by the field theory and the del Pezzo. This symmetry is a $\Z_2$ in 
the case of the second del Pezzo and exchanges nodes $4$ and $5$ in the quiver, 
or $E_1$ and $E_2$ in the geometry. In general, the $n$-th del Pezzo carries
the action of the Weyl group of $E_n$ and the quiver theory has the same
symmetries. Because the intersection form is invariant under the Weyl
group, one is left with a Weyl group ambiguity in identifying charges with
divisors. Our prescription using the dual exceptional collection fixes this 
ambiguity, and is only invariant under the simultaneous action on field 
theory and geometry.

\section{Three block collections, anomalies, and Markov}

Three block exceptional 
collections provide some of the most impressive evidence
in support of our prescription for identifying bifundamental fields
$X_{ij}$
with fractional divisors $D_{ij}$ in the dual geometry.
Our plan in this section is to present first a gauge theory
computation or description 
and then to relate the gauge theory to a property
of the exceptional collection.

\begin{figure}
\begin{center}
\psfrag{a }{$a$}
\psfrag{b }{$b$}
\psfrag{ c}{$c$}
\psfrag{1}{$1$}
\psfrag{2}{$2$}
\psfrag{SU(xN)}{$\SU(xN)$}
\psfrag{SU(yN)}{$\SU(yN)$}
\psfrag{SU(zN)}{$\SU(zN)$}
\psfrag{E}{$\cale$}
\psfrag{F}{$\calf$}
\psfrag{G}{$\calg$}
\epsfig{width=4.5in,file=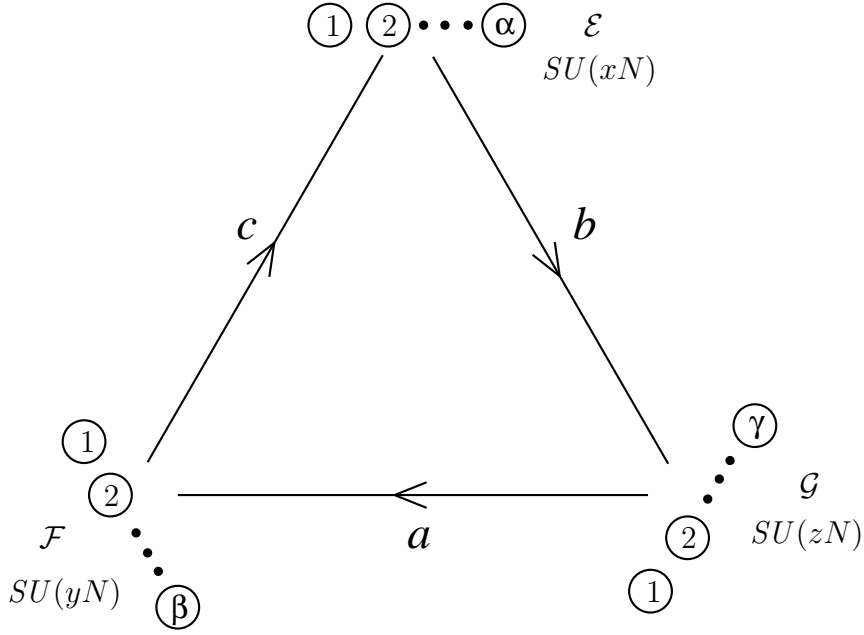}
\end{center}
\caption{Quiver for a three-block exceptional collection.}
\label{fig3block}
\end{figure}

Consider a quiver consisting of three blocks of nodes
(see Fig. \ref{fig3block}).  
The nodes within a block are not joined by arrows.  
The gauge groups of the nodes in a block are all the same.
Between
any two representative nodes in two distinct blocks, there
are the same number of arrows.  Let there be $\alpha$
nodes in the first block, $\beta$ nodes in the second block,
and $\gamma$ nodes in the third block.  Let the ranks of the
gauge groups be $xN$, $yN$, and $zN$.  Let there be $a$
arrows between nodes in the second and third blocks, $b$
arrows between nodes in the third and first blocks, and 
$c$ arrows between nodes in the first and second block.

These quivers derive from three block exceptional
collections $(\cale, \calf, \calg)$.  
For any two sheaves
within a block $E_i, \,  E_j \in \cale$, $\chi(E_i, E_j)=0$.
Moreover, for two sheaves in different blocks
$E_i \in \cale$ and $F_j \in \calf$, $\chi(E_i, F_j)$ is
independent of $i$ and $j$.  These three block 
collections satisfy in addition all the properties of
ordinary exceptional collections, and were 
discussed in great detail in \cite{wijn, NK}.
Three block exceptional collections
exist for $\CP{2}$, $\CP{1} \times \CP{1}$, and
all $dP_n$ for $n > 2$.

To see that the ranks of the gauge groups within a block
are all the same from $(\cale, \calf, \calg)$ 
requires more effort.  The requirement
$\chi(E_i, E_j)=0$ means that $r(E_i)/\deg(E_i)$ is 
independent of $i$.  One can then argue on more general
grounds that the rank and degree are coprime \cite{KO}.
The statement about the gauge groups 
follows from the fact that the $\chi$-dual
of $(\cale, \calf, \calg)$ is also a three block exceptional collection,
$(\calg \otimes K, L^D_\cale \calf, \cale)$.

We now analyze the gauge theory for these three block quivers.
The cancellation of the chiral anomaly requires, roughly
speaking, that the number of arrows into a node equals the
number of arrows out from that node.  More precisely,
\be
a \beta y = b \alpha x \; ; \; \; \;
b \gamma z = c \beta y \; ; \; \; \;
c \alpha x = a \gamma z \ .
\ee
These relations allow us to set $a$, $b$ and $c$ in terms
of the other variables up to a constant of proportionality $K'$.
In particular, we choose
\be
a = \alpha K' x \; ; \; \; \;
b = \beta K' y \; ; \; \; \;
c = \gamma K' z \ .
\ee
Knowing where this calculation is headed, we choose
\be
K' \equiv \sqrt{\frac{K^2}{\alpha \beta \gamma}} \ .
\ee

We want our gauge theory to be conformal, and thus we need the 
NSVZ beta functions to vanish for each node:
\begin{eqnarray}
x + \frac{1}{2} (b\gamma z (R_b - 1) + c \beta y (R_c - 1)) &=& 0  \ , \\
y + \frac{1}{2} (c \alpha x (R_c - 1) + a \gamma z (R_a -1)) &=& 0 \ , \\
z + \frac{1}{2} (a \beta y (R_a - 1) + b \alpha x (R_b - 1)) &=& 0 \ ,
\end{eqnarray}
where $R_a$, $R_b$, and $R_c$ are the R-charges of the bifundamental
matter fields.

We also insist that the superpotential, which we assume to be cubic,
have R-charge two.  This constraint $R_a + R_b + R_c = 2$ implies
the following Markov type equation for the quiver
\be
\frac{a^2}{\alpha} + \frac{b^2}{\beta} + \frac{c^2}{\gamma} = abc \ .
\label{Markovb}
\ee
We can rewrite this equation in terms of $x$, $y$, and $z$,
\be
\alpha x^2 + \beta y^2 + \gamma z^2 = \sqrt{K^2 \alpha \beta \gamma} xyz \ .
\label{Markov}
\ee

This Markov equation featured prominently in \cite{NK} where it was used
to classify all three block exceptional collections.  Geometrically,
this equation can be understood as an invariant of the matrix $S_{ij} = \chi(E_i, E_j)$.
The trace $\tr S^{-1} S^t$ is invariant under change of basis.  In the
special basis where the sheaves are written in terms of their rank, $c_1$,
and $\ch_2$, it is easy to check that $\tr S^{-1}S^t = 12 - K^2 = c_2(\bV)$.
In the basis where the sheaves form a three block exceptional
collection, $S$ takes the form
\be
S = \left(
\begin{array}{ccc}
\mathrm{Id}_\alpha & -C & B \\
0 & \mathrm{Id}_\beta & -A \\
0 & 0 & \mathrm{Id}_\gamma 
\end{array}
\right)
\ee 
where $\mathrm{Id}_n$ is an $n\times n$ identity matrix.
Also, $A$, $B$, and $C$ are rectangular matrices where every
entry is respectively $a$, $b$, or $c$.  
In order for
$\tr S^{-1} S^t$ to remain invariant, the Markov equation
(\ref{Markovb}) must hold.

Continuing with the gauge theory analysis, 
we find simple expressions for the R-charges
\be
R_a = \frac{2a}{\alpha bc} \; ; \; \; \;
R_b = \frac{2b}{\beta ac} \; ; \; \; \;
R_c = \frac{2c}{\gamma ab} \ .
\label{Rcharges}
\ee
These values agree with the geometric result (\ref{rcharge}).

We now consider the R-charges of dibaryonic operators in the gauge
theory.  Assume such an operator is constructed from $mN$ $a$-type
bifundamentals, $nN$ $b$-type bifundamentals, and $pN$ $c$-type bifundamentals.
The total R-charge is thus
\be
R/N = m R_a + n R_b + p R_c \ .
\ee
Using the expressions (\ref{Rcharges}), we can rewrite the total
R-charge as
\be
R/N = \frac{2}{K^2} \frac{1}{xyz} (max + nby + pcz) \ .
\ee

The dibaryons are constructed from antisymmetrizing over the 
fundamental indices of the matter fields.  To be able to
antisymmetrize, we need $x|(n-p)$, $y|(m-p)$, and $z|(m-n)$.
We also take the $x$, $y$, and $z$ to be relatively prime.
These conditions, along with the fact (which is a simple
consequence of (\ref{Markov})) that
\be
ax + by + cz \equiv 0 \mod xyz
\ee
imply that the total R-charge of the dibaryon is an integer multiple of
$2/K^2$ in agreement with
the geometric prediction (\ref{dibaryonr}).

We move on to a calculation of the cubic anomalies for these theories.
First we consider the $\tr R^3$ anomaly.
\begin{eqnarray}
\frac{1}{N^2} \tr R^3 &=& \alpha x^2 + \beta y^2 + \gamma z^2 \nonumber \\
&&
+ \beta \gamma yz a (R_a - 1)^3 + \alpha \gamma xz b (R_b - 1)^3
+ \alpha \beta xy c (R_c - 1)^3 \ .
\end{eqnarray}
Using (\ref{Markov}), this sum can be reduced to
\be
\frac{1}{N^2} \tr R^3 = \frac{24}{K^2} \ .
\label{trr3}
\ee

This result agrees with an independent prediction for the 
conformal anomaly $a_c$ in terms of the volume of
the Sasaki-Einstein manifold $\Y$.\footnote{
We add a subscript to the anomalies $a_c$ and $c_c$ 
in this section to
avoid confusion with the Euler characters $a$, $b$, and $c$.}   
From (\ref{a}),
and the fact that $\tr R = 0$, we know that 
$\tr R^3 = 32a_c/9$.  The vanishing of $\tr R$ also means that
$a_c=c_c$ where $c_c$ is the other central charge in the
superconformal algebra \cite{anselmi1, anselmi2}.  
The value of $c_c$ for ${\S^5}$ was
calculated long ago by \cite{gkp} to be $N^2/4$ and is easy
to generalize to arbitrary $\Y$ (knowing that $c_c$ is
proportional to $1/\Vol(\Y)$ \cite{skenderis, gubser}):
\be
a_c = c_c = \frac{N^2}{4} \frac{\Vol({\S^5})}{\Vol(\Y)} \ .
\label{aresult}
\ee  
For the del Pezzos, 
\be
\Vol(\Y)=\frac{\pi^3}{27} K^2 \ 
\label{voly}
\ee
as can be seen from, for example, \cite{BH}.  Putting
(\ref{aresult}), (\ref{voly}), and (\ref{a}) together, we
get precisely (\ref{trr3}).

There are also flavor $U(1)$ symmetries. It is easy to describe the
baryonic $\U(1)$'s discussed in subsections \ref{anomalies} and
\ref{intersec}. We order the nodes in each block in a line and label
them by their block and position. We find a baryonic $\U(1)$ for
each pair of adjacent nodes in a block as follows. An ingoing
bifundamental on the first node and an outgoing bifundamental
on the second node will have charge 1 under this $U(1)$.
An outgoing bifundamental on the second node and an ingoing
bifundamental on the first node will have charge $-1$ under
the $U(1)$.  We label the symmetry currents associated with 
these $U(1)$'s as $Q_I$, where $I$ runs over all pairs. 
There
are $\alpha-1$ such pairs from $\cale$, $\beta-1$ from $\calf$ and
$\gamma-1$ from $\calg$, for a total of $n$ baryonic $\U(1)$'s.
By construction, $\tr Q_I = 0$. It is straightforward to verify that
$\tr R^2 Q_I = 0$ and that $\tr Q_I Q_J Q_K = 0$. 

We can also investigate the properties of 
$\tr R Q_I Q_J$.  For convenience, we normalize the $U(1)$ charges  
by dividing out by the rank of the corresponding gauge
groups.   In particular, bifundamentals with charge
$\pm 1$ under a U(1) in the first block will now have charge
$\pm 1/xN$, bifundamentals with charge $\pm 1$ under
a U(1) in the second block will now have charge 
$\pm 1/yN$,
and bifundamentals with charge $\pm 1$ under a U(1)
in the third block will now have charge 
$\pm 1/zN$.
Using the NSVZ beta functions,
one finds that $\tr R Q_J^2 = -4$.  Moreover, for $Q_I$ and
$Q_J$ in the same block $\tr R Q_I Q_J = 2$, if $I$ and
$J$ have a node in common.  Otherwise, the trace vanishes.
One important point is that the matrix $\tr R Q_I Q_J$ has
all negative eigenvalues.  From \cite{IW2}, we conclude
that our choice of R symmetry maximizes $\tr R^3$ over the
space of $Q_I$.

It is amusing to see that the Dynkin diagrams of $E_n$ emerge in a
natural way out of these three block collections by using the bases of
$Q_I$ and their intersection given in the previous paragraph. Recall
that the $Q_I$ span the orthogonal complement of the R-charge in the
charge lattice, which is known to be isomorphic to the root lattice of
the exceptional Lie algebras.\footnote{We recall that by definition,
we have $E_5\cong D_5$, $E_4\cong A_4$, and $E_3\cong A_2\oplus A_1$.}
We can make this isomorphism very explicit for the three block
collections. To this end, consider the extended Dynkin diagrams of
$E_n$, and search for nodes such that the removal of this node leads
to diagrams consisting of (at most) three disconnected $A$-type Dynkin
diagrams, \ie, single lines of nodes. The resulting nodes correspond
to a basis of the root lattice of $E_n$. These bases can be mapped
onto the bases of $Q_I$ obtained from three-block collections, which
intersect in the same pattern. If one considers all possibilities of
removing one node in this way, one recovers exactly the lists of
integers $(\alpha,\beta,\gamma)$ that characterize the Markov
equations of Karpov and Nogin.  

\begin{figure}[t]
\begin{center}
\psfrag{E8}{$E_8$}
\psfrag{1}{$1$}
\psfrag{2}{$2$}
\psfrag{3}{$3$}
\psfrag{4}{$4$}
\psfrag{5}{$5$}
\psfrag{6}{$6$}
\psfrag{7}{$7$}
\psfrag{8}{$8$}
\psfrag{9}{$9$}
\epsfig{width=4in,file=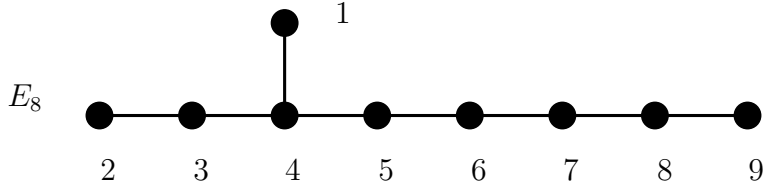}
\end{center}
\caption{Extended Dynkin diagram of $E_8$.}
\label{dynks}
\end{figure}

For example, consider the extended Dynkin diagram of $E_8$ in
Fig. \ref{dynks}. Removing node number $4$ leads to three groups of
nodes with $1$, $2$, and $5$ nodes in each group respectively. This means
that $(\alpha,\beta,\gamma)=(2,3,6)$ and corresponds to equation
number (8.3) on the list of \cite{NK}. Similarly, removing node $1$
leads to $(\alpha,\beta,\gamma)=(1,1,9)$ which is eq.\ (8.1), removing
node $3$ gives $(1,2,8)$, or eq.\ (8.2), and, finally, removing node
$5$ gives $(1,5,5)$ or eq.\ (8.4). In checking the similar statements
for the other del Pezzos one has to be careful that where one adds the
extending node to the Dynkin diagram of $E_n$ depends on $n$.
For $E_3$, one has to remove two nodes because there are two
extending nodes.

\section{The Eighth del Pezzo}

We crown this paper by applying our technology to the last del Pezzo surface
and all its 240 smallest dibaryons.
\begin{figure}
\begin{center}
\psfrag{1}{$1$}
\psfrag{2}{$2$}
\psfrag{8}{$8$}
\psfrag{9}{$9$}
\psfrag{10}{$10$}
\psfrag{11}{$11$}
\psfrag{SU(2N)}{$\SU(2N)$}
\psfrag{SU(N)}{$\SU(N)$}
\psfrag{SU(4N)}{$\SU(4N)$}
\psfrag{X}{$X$}
\psfrag{Y}{$Y$}
\psfrag{Z}{$Z$}
\epsfig{width=4.5in,file=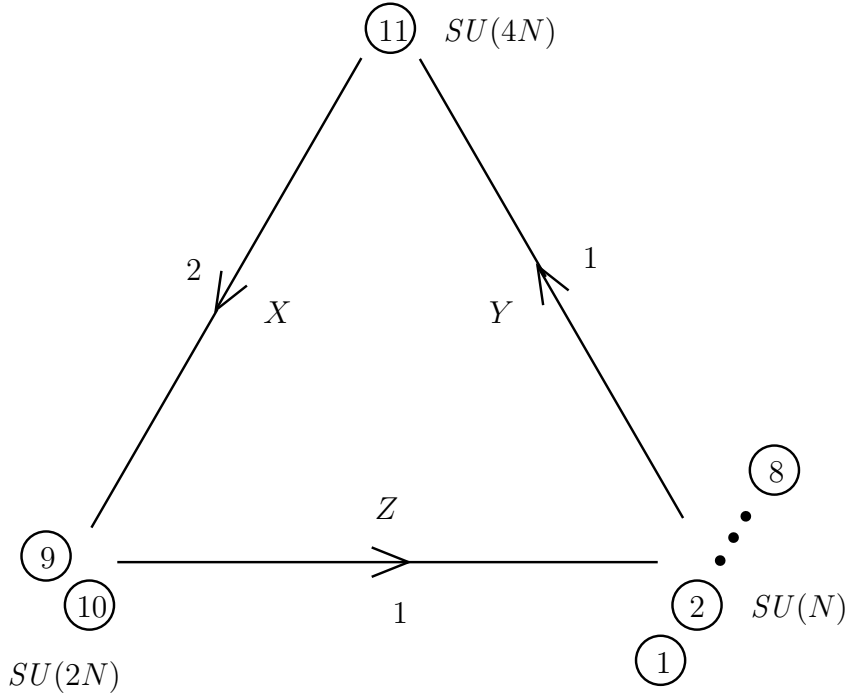}
\end{center}
\caption{Quiver for $dP_8$.}
\label{figdp8}
\end{figure}

To make the job easier, we pick the simplest dual three block collection 
$({\mathcal G}^\vee, {\mathcal F}^\vee, {\mathcal E}^\vee)$
that we know about.  In particular,
we choose the collection labeled (8.2) in the paper by Karpov and Nogin \cite{NK}.
The first block in the collection contains one node 
${\mathcal G}^\vee = (A)$ where $A$ is the right mutation of the trivial
bundle $\mO$ over the block of 
exceptional divisors $\mO(E_i)$, $i=4,\ldots,8$.  In particular,
$A$ falls into the short exact sequence
\be
0 \to \mO \to \bigoplus_{i=4}^8 \mO(E_i) \to A \to 0 \ .
\ee
Thus $\ch(A) = (4, \sum_{i=4}^8 E_i, -5/2)$.

The second block has two nodes ${\mathcal F}^\vee = (T, T')$ where 
$T$ and $T'$ are obtained through left mutations.  In particular,
$T'$ and $T$ are the left mutations of $\mO(H)$ 
and  $\mO(2H-E_1-E_2 - E_3)$ respectively over $\mO(H - E_i)$, $i=1,2,3$:
\be
0 \to 
\left\{ \begin{array}{c}
T' \\
T
\end{array} \right\}
\to \bigoplus_{i=1}^3 \mO(H-E_i) \to
\left\{ \begin{array}{c}
\mO(H) \\
\mO(2H-E_1-E_2-E_3)
\end{array}
\right\}
\to 0 \ .
\ee 
The charges are $\ch(T)=(2, H, -1/2)$ and $\ch(T')=(2,2H-E_1-E_2-E_3,-1/2)$.
The third block contains only line bundles.  In particular
\be
{\mathcal E}^\vee = \left( \left\{ \mO(E_i-K) \right\}_{i=4,\ldots8},
\left\{ \mO(H-E_j)\right\}_{j=1,2,3} \right) \ .
\ee

We label the bifundamental fields opposite from the third, second, and first
blocks $X$, $Y$, and $Z$ respectively.  
From either the gauge theory computation (\ref{Rcharges}) or the
geometric computation (\ref{rcharge}), we know the 
R-charges of these fields are
$R_X = 1/2$, $R_Y = 1/2$, and $R_Z = 1$.
In the interest of keeping things as simple as possible for this complicated
example, we restrict attention to the smallest dibaryons, \ie 
the dibaryons
with R-charge $2N$ or equivalently curves in the del Pezzo with degree one.

There are a number of different possible ways to construct these smallest
dibaryons from gauge theory.  
We go through each possibility and count up the naive number,
ignoring classical relations from the superpotential and any possible
``quantum'' relations (see \cite{Beasley1}).  We will see that we get too many
dibaryons.  We then repeat the calculation geometrically and see which
dibaryons we over counted and why.

There are naively three 
dibaryons constructed from antisymmetrizing over $4N$ copies
of the $X$ fields.  We can choose one of the $SU(2N)$ gauge groups
twice or each $SU(2N)$ gauge group once.  Geometrically,
the divisors for these three dibaryons are
\begin{eqnarray*}
2c_1(T) - c_1(A) &=&  2H - \sum_{i=4}^8 E_i \ , \\
2c_1(T') - c_1(A) &=& 4H - 2E_1-2E_2-2E_3 - \sum_{i=4}^8 E_i \ , \\
c_1(T) + c_1(T') - c_1(A) &=& 3H - \sum_{i=1}^8 E_i = -K \ . \\
\end{eqnarray*}
The first two divisors have genus zero, as can be seen from the
genus formula $K\cdot(K+C) = 2g-2$.  The last divisor, which
is nothing but the anticanonical bundle, has genus one.

We move onto the dibaryons constructed from antisymmetrizing over
$N$ copies of the $YXZ$ fields.  As there are eight nodes in the
last block, there are 64 possible dibaryons.  Geometrically,
to construct the divisor, we take the difference of 
two of the nodes and add $-K$:
\be
D_{YXZ} = c_1(E^\vee_{h(YXZ)}) - c_1(E^\vee_{t(YXZ)}) - K \ .
\ee
We see immediately that in the case that $t=h$, $D_{YXZ} = -K$.
Thus eight of these divisors are identical and 
have genus one.  So the gauge theory, without additional
relations, over counts by seven.  The remaining
56 divisors have genus zero, are all distinct,
and are summarized in (\ref{tabledp8}).  
Only the genus zero curves are tabulated.

Next we consider the dibaryons, constructed
from antisymmetrizing over $2N$ $Z$ type 
bifundamentals.  There are 56 such objects.
One might have thought it possible to antisymmetrize
over two copies of the same $SU(N)$ gauge group.  However,
since there is only one $Z$ type bifundamental between
any two nodes in the ${\mathcal F}^\vee$ and 
${\mathcal E}^\vee$ blocks, such
a dibaryon antisymmetrizes to give zero.  Thus,
we need to make sure the $SU(N)$ gauge groups are
distinct.  

Geometrically, we find exactly 56 $Z$ type divisors
with genus zero and degree one, as
tabulated in (\ref{tabledp8}).  One can ask what the 
analog of taking two identical $SU(N)$ gauge groups is.  
For example, consider the divisor
\be
2(H-E_1) - H = H - 2E_1 \ 
\ee  
obtained by taking two copies of $\mO(H-E_1)$ in the third
block.  This divisor does indeed have degree one, leading
to a dibaryon with R-charge $2N$.  However, the genus
is $-1$!  We conclude there is no such cycle for a D3-brane 
to wrap.  

There are 70 dibaryons formed from antisymmetrizing over
$4N$ copies of the $Y$ type bifundamentals.  Similar
to the $Z$ case, we need to make sure that all the 
$SU(N)$ gauge groups are distinct.
Geometrically
we find 70 distinct genus zero, degree one curves,
formed by taking the difference between $c_1(A)$ and
the first Chern classes of four distinct bundles in 
${\mathcal E}^\vee$ (plus $-K$).
The results are tabulated in (\ref{tabledp8}).

Finally, we consider the dibaryons of type $YX$.
Geometrically, the divisors are related to the divisors
of the $Z$ type dibaryons via $D \to D' = -2K - D$. 
It is easy to check that if the degree of $D$ is one,
then the degree of $D'$ is also one and moreover
that $D$ and $D'$ have the same genus.  
With a little more work to make sure that the $D'$
are distinct from the $D$, we find
the 56 additional degree one, genus zero dibaryons
in table \ref{tabledp8}.
These dibaryons are obtained by antisymmetrizing
over $2N$ copies of the $YX$ fields, making sure that
the antisymmetrization is over distinct $SU(N)$ gauge fields.
The requirement on the gauge theory side that the $SU(N)$
groups be distinct is presumably enforced by the
superpotential or additional ``quantum'' relations.
The argument provided for the $Z$ type dibaryons fails here.

Without a superpotential or knowledge of the additional
quantum relations we would have over counted the $YXZ$ and
$YX$ type dibaryons.  However, by identifying the bifundamental
fields with fractional divisors, it becomes clear which
gauge invariant combinations of the bifundamental fields 
correspond to which holomorphic curves.

\be
\renewcommand{\arraystretch}{1.2}
\begin{array}{r|cccccc}
& \mathrm{number} & X & Y & YX & Z & YXZ \\
E_i & 8 & 0 & 5 & 0 & 3 & 0\\
H-E_i-E_j & 28 & 0 & 0 & 10 & 3 & 15 \\
2H - \sum_{i=1}^5 E_{a_i} & 56 & 1 & 30 & 10 & 15 & 0 \\
3H - 2E_a - \sum_{i=1}^6 E_{a_i} & 56 & 0 & 0 & 15 & 15 & 26 \\
4H - 2E_a - 2E_b - 2E_c - \sum_{i=1}^5 E_{a_i} & 
56 & 1 & 30 & 15 & 10 & 0 \\
5H - 2\sum_{i=1}^6 E_{a_i} - \sum_{j=1}^2 E_{a_j} & 
28 & 0 & 0 & 3 & 10 & 15\\
6H - 3E_a - 2\sum_{i=1}^7 E_{a_i} & 8 & 0 & 5 & 3 & 0 & 0 \\
\hline
\mathrm{totals} & 240 & 2 & 70 & 56 & 56 & 56
\end{array}
\label{tabledp8}
\ee

\begin{acknowledgments} 
It is a pleasure to thank James McKernan, Andrei Mikhailov, 
Radu Roiban, Cumrun Vafa, and Brian 
Wecht for useful discussions and illuminating comments. We are grateful to Martijn 
Wijnholt for helpful correspondence. This work is supported in part by the NSF under 
Grant No.\ PHY99-07949. 
\end{acknowledgments}




\providecommand{\href}[2]{#2}\begingroup\raggedright 
\endgroup


\begin{thebibliography}{1} 


\bibitem{jthroat}
J.~Maldacena, ``The Large N limit of superconformal field theories and
supergravity,'' {\it Adv. Theor. Math. Phys.} {\bf 2} (1998) 231, 
{{\tt hep-th/9711200}}.

\bibitem{gkp}
S.S. Gubser, I.R. Klebanov, and A.M. Polyakov, ``Gauge theory correlators
from noncritical string theory,''
{\it Phys. Lett.} {\bf B428} (1998) 105,
{{\tt hep-th/9802109}}.

\bibitem{EW}
E.~Witten, ``Anti-de Sitter space and holography,''
{\it Adv. Theor. Math. Phys.} {\bf 2} (1998) 253,
{{\tt hep-th/9802150}}.

\bibitem{aharony}
O.~Aharony,
``The non-AdS/non-CFT correspondence, or three different paths to QCD,''
{\tt hep-th/0212193};
F.~Bigazzi, A.~L.~Cotrone, M.~Petrini and A.~Zaffaroni,
``Supergravity duals of supersymmetric four dimensional gauge theories,''
{\tt hep-th/0303191}.

\bibitem{LNV}
A.~Lawrence, N.~Nekrasov, and C.~Vafa, ``On conformal field theories in 
four-dimensions,'' {\it Nucl. Phys.} {\bf B533} (1998) 199,
{\tt hep-th/9803015}.

\bibitem{KS}
S.~Kachru and E.~Silverstein, ``4-D conformal theories and strings on orbifolds,''
{\it Phys. Rev. Lett.} {\bf 80} (1998) 4855, 
{\tt hep-th/9802183}.

\bi{Kehag}
A. Kehagias, ``New Type IIB Vacua and Their F-Theory Interpretation,''
{\it Phys. Lett.}  {\bf B435} (1998) 337, 
{{\tt hep-th/9805131}}.

\bi{Acharya}
B.~S.~Acharya, J.~M.~Figueroa-O'Farrill, C.~M.~Hull and B.~Spence,
``Branes at conical singularities and holography,''
{\it Adv. Theor. Math. Phys.}  {\bf 2} (1999) 1249,
{\tt hep-th/9808014}.

\bi{MoPle}
D.~Morrison and R.~Plesser,
``Non-Spherical Horizons, I,''
{\it Adv. Theor. Math. Phys.} {\bf 3} (1999) 1,  {\tt hep-th/9810201}.

\bibitem{KW}
I.~R.~Klebanov and E.~Witten, ``Superconformal field theory on three-branes
at a Calabi-Yau singularity,'' 
{\it Nucl. Phys.} {\bf B536} (1998) 199, 
{\tt hep-th/9807080}.

\bibitem{klst}
I.~R.~Klebanov and M.~J.~Strassler,
``Supergravity and a confining gauge theory: Duality cascades and  chiSB-resolution of naked singularities,''
{\it JHEP} {\bf 0008} (2000) 052,
{\tt hep-th/0007191}.

\bi{Z3orb}
S.~Gukov, M.~Rangamani and E.~Witten,
``Dibaryons, strings, and branes in AdS orbifold models,''
{\it JHEP} {\bf 9812} (1998) 025,
{\tt hep-th/9811048}.

\bibitem{EW2}
E.~Witten,
``Baryons and branes in anti de Sitter space,''
{\it JHEP} {\bf 9807} (1998) 006,
{\tt hep-th/9805112}.

\bibitem{KG}
S.~S.~Gubser and I.~R.~Klebanov
``Baryons and Domain Walls in an
{\cal N}=1 Superconformal Gauge Theory,'' {\it Phys. Rev.} {\bf D58} 
(1998) 125025, {\tt hep-th/9808075}.

\bi{Beasley1}
C.~E.~Beasley and M.~Ronen~Plesser, 
``Toric Duality is Seiberg Duality,''
{\it JHEP} {\bf 0112} (2001) 001,
{\tt hep-th/0109053}.

\bi{HM}
C.~P.~Herzog and J.~McKernan,
``Dibaryon Spectroscopy,''
{\tt hep-th/0305048}.

\bibitem{IW2}
K.~Intriligator and B.~Wecht,
``Baryon charges in 4D superconformal field theories and their AdS  duals,''
{\tt hep-th/0305046}.

\bibitem{DM}
M.~Douglas and G.~Moore, ``D-Branes, Quivers and ALE Instantons,''
{\tt hep-th/9603167}.

\bibitem{beasley0}
C.~Beasley, B.~R.~Greene, C.~I.~Lazaroiu and M.~R.~Plesser,
``D3-branes on partial resolutions of abelian quotient singularities of  Calabi-Yau threefolds,''
{\ it Nucl. Phys.} {\bf B566} (2000) 599,
{\tt hep-th/9907186}.

\bibitem{fhh1}
B.~Feng, A.~Hanany and Y.~H.~He,
``D-brane gauge theories from toric singularities and toric duality,''
{\it Nucl. Phys.}  {\bf B595} (2001) 165,
{\tt hep-th/0003085}.

\bibitem{fhh2}
B.~Feng, A.~Hanany and Y.~H.~He,
``Phase structure of D-brane gauge theories and toric duality,''
{\it JHEP} {\bf 0108} (2001) 040,
{\tt hep-th/0104259}.

\bibitem{unify}
F.~Cachazo, B.~Fiol, K.~A.~Intriligator, S.~Katz and C.~Vafa,
``A geometric unification of dualities,''
{\it Nucl. Phys.} {\bf B 628} (2002) 3,
{\tt hep-th/0110028}.

\bibitem{digo}
D.~E.~Diaconescu and J.~Gomis,
``Fractional branes and boundary states in orbifold theories,''
{\tt JHEP} {\bf 0010} (2000) 001,
{\tt hep-th/9906242}.

\bibitem{DFR}
M.~R.~Douglas, B.~Fiol and C.~Romelsberger,
``The spectrum of BPS branes on a noncompact Calabi-Yau,''
{\tt hep-th/0003263}.

\bibitem{dido}
D.~E.~Diaconescu and M.~R.~Douglas,
``D-branes on stringy Calabi-Yau manifolds,''
{\tt hep-th/0006224}.

\bibitem{mayr}
P.~Mayr,
``Phases of supersymmetric D-branes on Kaehler manifolds and the McKay correspondence,''
{\tt JHEP} {\bf 0101}, 018 (2001) 018,
{\tt hep-th/0010223}.

\bibitem{douglas}
M.~R.~Douglas,
``D-branes, categories and N = 1 supersymmetry,''
{J.  Math.  Phys.} {\bf 42} (2001) 2818,
{\tt hep-th/0011017}.

\bibitem{trieste}
M.~R.~Douglas,
``Lectures on D-branes on Calabi-Yau manifolds,''
\href{http://www.slac.stanford.edu/spires/find/hep/www?irn=5015596}{SPIRES entry}
{\it Prepared for ICTP Spring School on Superstrings and Related Matters, Trieste, Italy, 2-10 Apr 2001}.

\bibitem{zaslow}
E.~Zaslow,
``Solitons and helices: The Search for a math physics bridge,''
{\it Commun. Math. Phys.}  {\bf 175} (1996) 337,
{\tt hep-th/9408133}.

\bibitem{hiv}
K.~Hori, A.~Iqbal and C.~Vafa,
``D-branes and mirror symmetry,''
{\tt hep-th/0005247}.

\bibitem{haiq}
A.~Hanany and A.~Iqbal, 
``Quiver Theories from D6-branes via Mirror Symmetry,''
{\it JHEP} {\bf 0204} (2002) 009,
{\tt hep-th/0108137}.

\bibitem{ceva}
S.~Cecotti and C.~Vafa,
``On classification of N=2 supersymmetric theories,''
{\it Commun. Math. Phys.}  {\bf 158} (1993) 569,
{\tt hep-th/9211097}.

\bibitem{wijn}
M.~Wijnholt,
``Large volume perspective on branes at singularities,''
{\tt hep-th/0212021}.

\bibitem{NK}
B. V. Karpov\ and\ D. Yu. Nogin, 
``Three-block Exceptional Collections over del Pezzo Surfaces,''
Izv. Ross. Akad. Nauk Ser. Mat. {\bf 62} (1998), no. 3, 3--38; 
translation in Izv. Math. {\bf 62} (1998), no.~3, 429--463,
{\tt alg-geom/9703027}.

\bibitem{wall}
A.~Hanany and J.~Walcher,
``On duality walls in string theory,''
{\tt hep-th/0301231}.

\bibitem{IW}
K.~Intriligator and B.~Wecht,
``The exact superconformal R-symmetry maximizes $a$,''
{\tt hep-th/0304128}.

\bibitem{fhh3}
B.~Feng, A.~Hanany, Y.~H.~He and A.~Iqbal,
``Quiver theories, soliton spectra and Picard-Lefschetz transformations,''
{\it JHEP} {\bf 0302} (2003) 056,
{\tt hep-th/0206152}.

\bibitem{bedo}
D.~Berenstein and M.~R.~Douglas,
``Seiberg duality for quiver gauge theories,''
{\tt hep-th/0207027}.

\bibitem{braun}
V.~Braun,
``On Berenstein-Douglas-Seiberg duality,''
{\it JHEP} {\bf 0301} (2003) 082,
{\tt hep-th/0211173}.

\bibitem{GNS}
S.~G.~Gubser, N.~Nekrasov, and S.~Shatashvili, ``Generalized Conifolds and Four 
Dimensional ${\cal N}=1$ Superconformal Theories,'' 
{\it JHEP} {\bf 9905} (1999) 003,
{\tt hep-th/9811230}.

\bibitem{rudakov}
``Helices and vector bundles,''
Seminaire Rudakov. London Mathematical Society Lecture Note Series, 148. 
Cambridge University Press, Cambridge, 1990.

\bibitem{KO}
S. A. Kuleshov\ and\ D. O. Orlov, 
``Exceptional sheaves over del Pezzo surfaces,''
Izv. Ross. Akad. Nauk Ser. Mat. {\bf 58} (1994), no. 3, 53--87; 
translation in Russian Acad. Sci. Izv. Math. {\bf 44} (1995), no.~3, 479--513.

\bibitem{oova}
H.~Ooguri and C.~Vafa,
``Geometry of N = 1 dualities in four dimensions,''
{\it Nucl. Phys.} {\bf B500} (1997) 62,
{\tt hep-th/9702180}.

\bibitem{anselmi1}
D.~Anselmi, D.~Z.~Freedman, M.~T.~Grisaru and A.~A.~Johansen,
``Nonperturbative formulas for central functions of supersymmetric gauge  theories,''
{\it Nucl. Phys.} {\bf B526} (1998) 543,
{\tt hep-th/9708042}.

\bibitem{anselmi2}
D.~Anselmi, J.~Erlich, D.~Z.~Freedman and A.~A.~Johansen,
``Positivity constraints on anomalies in supersymmetric gauge theories,''
{\it Phys. Rev. D} {\bf 57} (1998) 7570,
{\tt hep-th/9711035}.

\bi{Beasley2}
C.~E.~Beasley,
``BPS Branes from Baryons,''
{\it JHEP} {\bf 0211} (2002) 015,
{\tt hep-th/0207125}.

\bi{Mikhailov}
A.~Mikhailov,
``Giant gravitons from holomorphic surfaces,''
{\it JHEP}  {\bf 0011} (2000) 027,
{\tt hep-th/0010206}.

\bibitem{BHK}
D.~Berenstein, C.~P.~Herzog and I.~R.~Klebanov,
``Baryon spectra and AdS/CFT correspondence,''
{\it JHEP} {\bf 0206} (2002) 047,
{\tt hep-th/0202150}.

\bibitem{mysterious}
A.~Iqbal, A.~Neitzke and C.~Vafa,
``A mysterious duality,''
{\it Adv. Theor. Math. Phys.}  {\bf 5} (2002) 769,
{\tt hep-th/0111068}.

\bibitem{dofi}
M.~R.~Douglas and B.~Fiol,
``D-branes and discrete torsion. II,''
{\tt hep-th/9903031}.

\bibitem{mrd}
M.~R.~Douglas, comments quoted in \cite{IW2}.

\bibitem{skenderis} {M.~Henningson and K.~Skenderis,
``The holographic Weyl anomaly,''
{\it JHEP} {\bf 9807} (1998) 023,
{\tt hep-th/9806087}.}

\bibitem{gubser}
S.~S.~Gubser, ``Einstein Manifolds and Conformal Field Theories,'' 
{\it Phys. Rev. D} {\bf 59} (1999) 025006, 
{\tt hep-th/9807164}.

\bibitem{BH}
A.~Bergman and C.~P.~Herzog,
``The volume of some non-spherical horizons and the AdS/CFT
correspondence,''
{\it JHEP} {\bf 0201} (2002) 030,
{\tt hep-th/0108020}.

\end{thebibliography}
\end{document}